\documentclass[10pt,twocolumn]{article} 
\usepackage{times}
\usepackage{graphicx}
\usepackage{amssymb}
\usepackage{url}
\usepackage{amsmath}
\usepackage{graphicx}
\usepackage{hyperref}
\usepackage{cite}
\usepackage{color}
\usepackage{xfrac}
\usepackage{epstopdf}
\usepackage{tabularx}
\usepackage{braket}
\usepackage[mathlines,switch]{lineno} %Add numbers to the text lines
\usepackage[normalem]{ulem}
\usepackage{xcolor}
\usepackage[bottom]{footmisc}

\usepackage{titlesec}

\titleformat*{\section}{\large\bfseries}
\titleformat*{\subsection}{\normalsize\bfseries}

\usepackage[scale=0.85]{geometry}
\newcommand{\lSAW}{{\lambda_{\mathrm{SAW}}}}	% acoustic wavelength
\newcommand{\fSAW}{f_\mathrm{SAW}}			% acoustic SAW frequency
\newcommand{\mycomment}[1]{}

\let\OLDthebibliography\thebibliography
\renewcommand\thebibliography[1]{
  \OLDthebibliography{#1}
  \setlength{\parskip}{0pt}
  \setlength{\itemsep}{0pt plus 0.6ex}
}

\begin{document}

\title{\Large{\textbf{Optical waveguide devices modulated by surface acoustic waves}}}

\author{Antonio Crespo Poveda$^{1,2,\ast}$, Dominik D. B\"{u}hler$^{1}$, Andr{\'e}s Cantarero S{\'a}ez$^{3}$ \\
Paulo V. Santos$^{2}$, and Maur{\'i}cio M. de Lima, Jr.$^{1,\dagger}$
\\
\vspace*{-0.2cm}
\\
\small{$^1$ Materials Science Institute Universitat de Val\`encia, PO Box 22085, 46071 Valencia (Spain)}\\
\small{$^2$ Paul-Drude-Institut f\"{u}r Festk\"{o}rperelektronik, Hausvogteiplatz 5-7, 10117 Berlin (Germany)}\\
\small{$^3$ Institute of Molecular Science Universitat de Val\`encia, PO Box 22085, 46071 Valencia (Spain)}\\
\medskip
\footnotesize{$^\ast$\textit{antonio.poveda@pdi-berlin.de} and $^\dagger$\textit{mmlimajr@uv.es}}}  

\date{}

\maketitle
\thispagestyle{empty}

\begin{abstract}

\noindent This paper reviews the application of coherent acoustic phonons in the form of piezoelectrically-generated surface acoustic waves to control the response of integrated photonic acousto-optic waveguide structures. We first address the fundamental properties of the acoustic fields in piezoelectric materials as well as the piezoelectric generation of surface acoustic waves using interdigital transducers. The mechanism responsible for the interaction between light and the acoustic modes in the photonic waveguide structures and the modulation of the response of the devices is carefully reviewed. Next, we discuss the most important developments in the field of integrated acousto-optical device applications, with focus on devices built upon three-dimensional optical waveguides. Finally, prospects for acousto-optic device applications by use of surface acoustic waves are summarized. 
  
\end{abstract}

\tableofcontents

\section{Introduction}\label{Introduction}

Most of nowadays long-distance information transmission is made in the optical domain by optical fibres arranged in large optical networks worldwide. Nevertheless, most of the transmitted information is still processed electronically at the nodes of the optical networks, which substantially limits data throughput, and the device level. This makes it necessary to develop efficient interfaces for the conversion of photons into electrons and vice versa, which is undesirable from a pragmatic perspective. Moreover, the achievable time response of electronic systems is considerably slower than the one expected for optical components. Therefore, in order to take advantage of the fast time response inherent to optical phenomena, it is also necessary to perform at least part of the information processing in the optical domain. 

The development of integrated photonic circuits for the effective optical processing of information requires the efficient confinement and guiding of light at the micron-scale using integrated optical waveguides (WGs), as well as methods to control its propagation. The latter can be achieved if the WG structure is turned into a tunable medium by changing its dimensions or optical properties as required. The most straightforward way of actively controlling the propagation of light in WGs consists in changing the temperature of the medium through the thermo-optic effect. Temperature changes are accompanied by changes in the volume of the medium (i.e. the atomic spacing in the crystalline structure), which directly effects its electro-optic properties. This results in the tuning of the dielectric constant $\epsilon$ and the refractive index $n$ of the WG material. Some examples of thermally-tuned devices reported so far with time responses in the low MHz frequency range include a Si Fabry-P\'{e}rot micro-cavity in which tuning is achieved by way of direct current flow through the structure~\cite{Pruessner:07}, or the direct heating of a Si WG by use of a p-n junction embedded in the WG center~\cite{Li:14}. A further strategy makes use of electric fields to change the optical properties of the WG material. If the latter lacks center of inversion (i.e., it is piezoelectric), the change of $n$ is in proportion to the applied electric field (linear electro-optic effect or Pockels effect). This tuning technique enables very fast devices operating in the GHz range~\cite{Yang:11,Xing:13}. If the WG material has inversion symmetry (i.e, it is non-piezoelectric), the change of the refractive index is in proportion to the square or even higher powers of the electric field (Kerr effect). In practical devices, this technique is usually implemented by irradiating the WG material with a secondary optical beam of high intensity. The electric field of the incident secondary light beam interferes with the underlying media, modulating its optical properties and enabling very high switching frequencies~\cite{Lu:11,doi:10.1021/nl404356t,almeida2004all}. Additionally, free-carrier injection by means of electric or optical means has been investigated for tunable photonic structures in GaAs/AlGaAs~\cite{1159054}, InP~\cite{1386333}, or Si~\cite{1073206}. In this case, the refractive index increases or decreases when carriers are respectively depleted from or injected into the WG material~\cite{1073206}. Finally, externally applied strain has also been proposed as a tuning technique for integrated photonics~\cite{doi:10.1063/1.1371786,doi:10.1063/1.1649803}. As will be explained in Section~\ref{Acousto-optical interaction in semiconductor structures}, the dielectric properties of the WG material are modified in this case by the variations of the volume and symmetry of the crystal induced by the strain field (elasto-optic effect).

In this paper, we review the modulation of the optical properties of photonic structures using the dynamic strain field generated by acoustic waves. In particular, we focus on acoustic modes that propagate close to the surface of the medium or at the interface between two materials with different elastic properties. These modes are generally referred as surface acoustic waves (SAWs). In piezoelectric materials, SAWs can be electrically generated by planar interdigital transducers (IDTs)~\cite{doi:10.1063/1.1754276} and offer, therefore, the best possibilities for active integrated photonic applications. Historically, the initial studies addressed the unguided propagation of  sound and light in bulk structures. The development of the first IDTs for the piezoelectric generation of SAWs and the first planar two-dimensional (2D) optical WGs~\cite{1126862} made it possible to scale acousto-optic functionalities down to the chip level. In this way, first integrated acousto-optic devices were mainly based on Bragg diffraction of guided light beams by SAW beams in 2D slab optical WGs. A whole class of integrated acousto-optical devices devices, such as light modulators~\cite{1076309,doi:10.1063/1.1654747} or light switches~\cite{doi:10.1063/1.106625,145254}, were demonstrated during the first years. An extensive review of the earliest developments can be found in e.g.~\cite{tsai2013guided} and references therein. In this topical review, we will concentrate on recent developments related to the acousto-optical modulation of three-dimensional (3D) rectangular WGs. The paper is organized as follows: Section~\ref{Overview of optical waveguiding} gives an overview of optical waveguiding using 3D rectangular WGs. Section~\ref{Surface acoustic waves in piezoelectric semiconductors} describes the properties of surface acoustic modes as well as the fundamentals of SAW excitation and guiding. The effects of the SAW-induced strain on the optical properties of semiconductors, as well as the modulation of the optical properties of WG structures are then described in Section~\ref{Acousto-optical interaction in semiconductor structures}. In the subsequent sections, we review the acoustic modulation of different types of photonic structures, namely ring resonators (Section~\ref{Ring resonators modulated by surface acoustic waves}) and interferometers and multiplexers (Section~\ref{Interferometric acousto-optic devices}). Finally, some conclusions of the state of the art as well as some futures perspectives are given in Section~\ref{Conclusions and perspectives}.

\section{Overview of optical waveguiding}\label{Overview of optical waveguiding}

%%%%%%%%%%%%%%%%%%%%%%%%%%%%%%%%%%%%%%%%%%%%%%%%%%%%%%%%%%%%%%%%%%%%%%%%%%%%%%%%%%%%%%
\begin{figure*}[t!]
\begin{center}
		\includegraphics[width=0.86\textwidth]{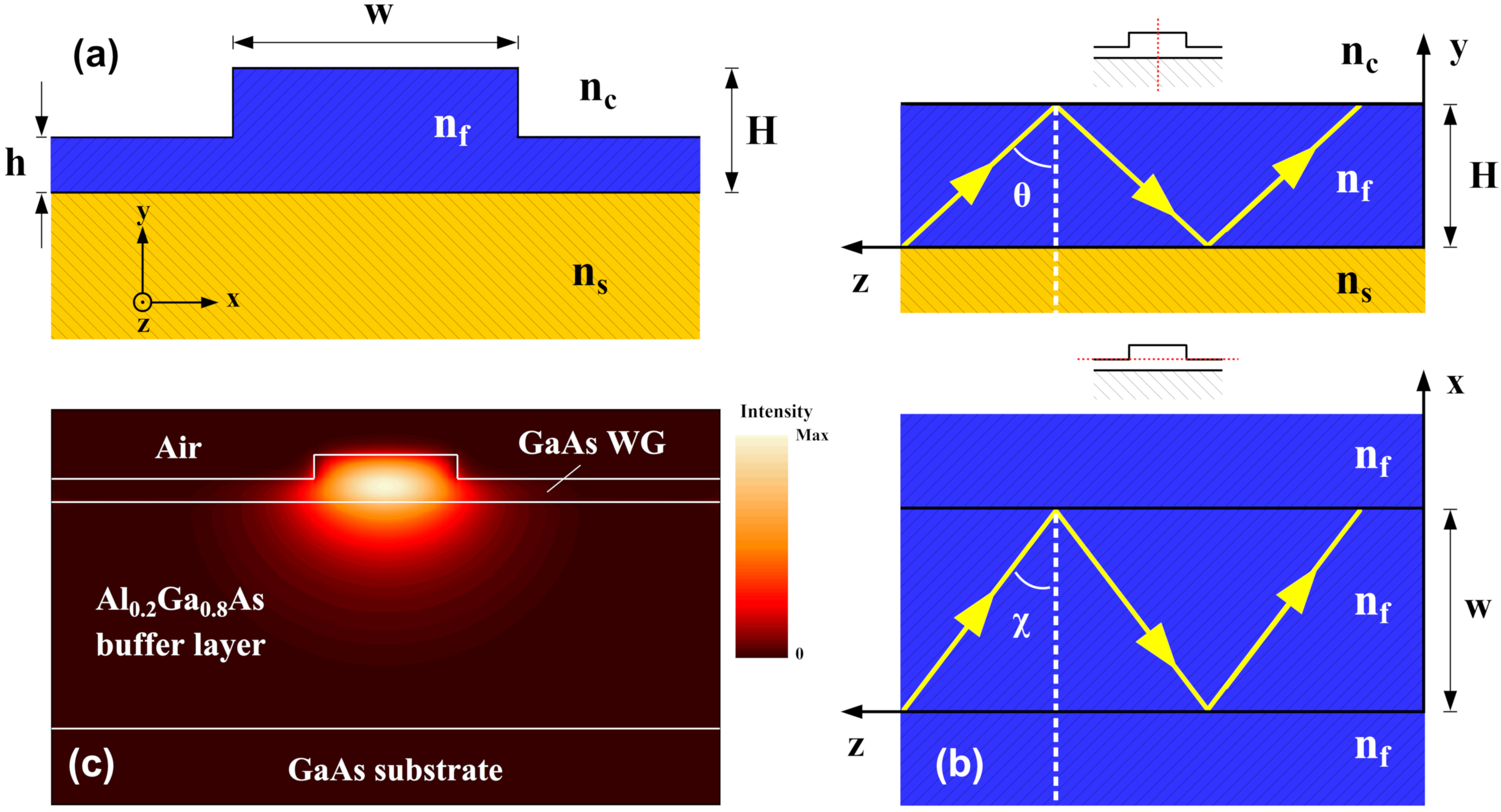}
	\caption{(a) Drawing of a 3D rectangular WG (also referred as \textit{rib} WG) of width $w$, etched to a depth $h$. The guiding layer, with refractive index $n_{f}$, is placed on top of a layer of refractive index $n_{s}$ and capped by a layer of refractive index $n_{c}$, with $n_{f}>\left(n_{s},n_{c}\right)$. (b) Side (upper panel) and top view (lower panel) of the WG structure shown in (a), together with the rays of the guided modes. When the angle of incidence reaches $\Theta$ (in the $x-z$ plane) and $\Xi$ (in the $y-z$ plane), total internal reflection occurs at the interfaces between WG regions with different refractive index. As a consequence, the optical waves travel following a zig-zag path along the rectangular WG section. (c) Optical field distribution corresponding to the fundamental $\mathrm{HE_{00}}$ mode of a GaAs/AlGaAs rib WG structure calculated for $\lambda=900$~nm, $h=150$~nm, $H=300$~nm, and $w=900$~nm.}
	\label{fig: WGs_composition}
\end{center}
\end{figure*}
%%%%%%%%%%%%%%%%%%%%%%%%%%%%%%%%%%%%%%%%%%%%%%%%%%%%%%%%%%%%%%%%%%%%%%%%%%%%%%%%%%%%%%

Integrated optical WGs play a fundamental role in the transmission of optical information between the different photonic components just like conductive pads in integrated electronic circuits. The simplest planar WGs are based on a semiconductor 2D planar layer surrounded by two cladding layers with lower refractive index~\cite{1126862}, which confine light via total internal reflection. By placing rectangular boundary walls, 3D rectangular WGs~\cite{Goell:73} can be engineered to provide light confinement in an additional spatial dimension. Here, the higher light confinement allows for minimizing the interaction (crosstalk) between adjacent components, thus enabling dense  on-chip integration of photonic structures. During the past forty years, the vast majority of developments such as optical modulators/switches~\cite{doi:10.1063/1.88279,doi:10.1063/1.87992,390234} or on-chip lasers~\cite{250348,849055}, among others, have made use of 3D rectangular WGs. 

Figure~\ref{fig: WGs_composition}(a) shows a sketch of a 3D rectangular WG (also referred as \textit{rib} WG) of width $w$. The structure is etched to a depth $H-h$ on a guiding layer of refractive index $n_{f}$ placed on top of a layer of refractive index $n_{s}$ and capped by a layer of refractive index $n_{c}$ (usually air), satisfying the conditions $n_{f}>\left(n_{s},n_{c}\right)$. If the guiding layer is completely etched, the structure is commonly referred as \textit{ridge} WG. The guiding mechanism of the WG is described in Figure~\ref{fig: WGs_composition}(b), where a side ($y-z$ plane, upper panel) and top view ($x-z$ plane, lower panel) of the WG structure, together with the rays of the guided modes, are displayed. The dotted red line in the diagram on top of each panel indicates the position of the view plane with respect to (a). In general, guided modes (i.e. field distributions that do not change their shape during propagation) can only propagate if total internal reflection occurs both at the top and lateral WG walls. Whether this occurs or not is determined by two critical angles, namely $\theta$ ($y-z$ plane) and $\chi$ ($x-z$ plane), which depend upon the refractive index contrast at the top and lateral interfaces, respectively. When the angle of incidence $\alpha>\left(\theta,\chi\right)$, total internal reflection occurs at both interfaces and the optical waves travel following a zig-zag path along the rectangular WG section. The optical fields that do not satisfy the total internal reflection conditions are not guided and couple to radiation or substrate modes. A thorough explanation as well as a complete mathematical treatment can be found e.g. in~\cite{ebeling1993integrated}.

The propagation of the optical fields in WGs of arbitrary shape is described by the vector wave equations for the electric $\vec{E}$ and magnetic $\vec{H}$ fields. These, assuming that the material forming the WGs is dielectric, and non-magnetic, can be written as~\cite{ebeling1993integrated,book:166107}:
\begin{eqnarray}
\nabla^{2}\vec{E}-\mu_{0}\epsilon_{0}n^{2}\frac{\partial^{2}\vec{E}}{\partial t^{2}}=0 \label{Maxwell_vector_eqs_a}\\
\nabla^{2}\vec{H}-\mu_{0}\epsilon_{0}n^{2}\frac{\partial^{2}\vec{H}}{\partial t^{2}}=0 \label{Maxwell_vector_eqs_b}
\end{eqnarray}
where $\vec{\nabla}\equiv(\partial/\partial x,\partial/\partial y,\partial/\partial z)$, $n$ is the refractive index, $\mu_{0}$ and $\epsilon_{0}$ are the vacuum permeability and permittivity, respectively, and $\times$ indicates the vector product. If we assume monochromatic fields with a sinusoidal time dependence, \ref{Maxwell_vector_eqs_a}--\ref{Maxwell_vector_eqs_b} can be rewritten as~\cite{ebeling1993integrated,book:166107}:  
\begin{eqnarray}
\nabla^{2}\vec{E}-\mu_{0}\epsilon_{0}n^{2}\omega^{2}\vec{E}=0 \label{Maxwell_vector_eqs_a_helmholtz}\\
\nabla^{2}\vec{H}-\mu_{0}\epsilon_{0}n^{2}\omega^{2}\vec{H}=0 \label{Maxwell_vector_eqs_b_helmholtz}
\end{eqnarray}
with $\omega$ the angular frequency. Assuming that the optical fields propagate along the $\hat{z}$ axis, \ref{Maxwell_vector_eqs_a_helmholtz}--\ref{Maxwell_vector_eqs_b_helmholtz} admit solutions of the form~\cite{ebeling1993integrated,book:166107}:
\begin{eqnarray}
\vec{E}=\vec{E}_{\nu}e^{-\mathit{i}\beta_{\nu}z} \label{Fields_modes_a}\\
\vec{H}=\vec{H}_{\nu}e^{-\mathit{i}\beta_{\nu}z}, \label{Fields_modes_b}
\end{eqnarray}
corresponding to the WG modes, with $\nu$ denoting the order of the mode. It should be noted that, in 3D WGs, $\nu$ stands for the pair $\left(m,p\right)$, with $m$ and $p$ the number of nodes along the $\hat{x}$ and $\hat{y}$ directions, respectively. Here, $z$ is the light propagation direction, $\vec{E}_{\nu}$ and $\vec{H}_{\nu}$ are the field modal distributions, and $\beta_{\nu}$ is the phase propagation constant of each mode, which describes the phase change of the propagating field along the WG. Thus, $\beta_{\nu}$ is a measure of the wave number $k=2\pi/\lambda_{}$ inside the material, $\lambda_{}$ being the light wavelength. The phase propagation constant allows us to define two important magnitudes, namely the effective refractive index $n_{\mathrm{eff},\nu}$ and the phase velocity $v$ of the WG mode $\nu$, as~\cite{ebeling1993integrated,book:166107}:  
\begin{equation}
\beta_{\nu}=n_{\mathrm{eff},\nu}\frac{\omega}{c}=\frac{\omega}{v}.
\label{Effective index}  
\end{equation}

Here, $n_{\mathrm{eff},\nu}$ quantifies the refractive index that is effectively ``seen" by the WG mode $\nu$ during propagation along complex structures with a given refractive index spatial distribution (as, for instance, 3D WGs), and therefore depends strongly on the geometry of the structure. It can be computed numerically using a number of techniques including e.g. the effective index method~\cite{ebeling1993integrated,book:166107}. 

All modes of a WG having a different propagation constant $\beta_{\nu}$ along the $\hat{z}$ direction are mutually orthogonal. It is useful to decompose the fields into the longitudinal field components $\vec{E}_{z}=\left(0,0,E_{z}\right)$ and $\vec{H}_{z}=\left(0,0,H_{z}\right)$, and into the transverse field components $\vec{E}_{t}=\left(E_{x},E_{y},0\right)$ and $\vec{H}_{t}=\left(H_{x},H_{y},0\right)$. This allows us to write the orthogonality relation between two modes with different propagation constants along the $\hat{z}$ axis, $\beta_{\nu}\neq\beta_{\mu}$, as~\cite{ebeling1993integrated}: 
\begin{equation}
\int\int_{-\infty}^{\infty}\vec{E}_{t,\nu}\times\vec{H}_{t,\mu}^{*}dxdy=0~~~~\mathrm{for}~\beta_{\nu}\neq\beta_{\mu}
\label{Orthogonality_relation}  
\end{equation}
where $*$ stands for the complex conjugate. A detailed theoretical demonstration of this important relation can be found in~\cite{ebeling1993integrated,book:166107}.  

In the simplest case of 2D planar WGs, the guided modes can be described in terms of TE (transverse electric, i.e. $H_{y}=E_{z}=E_{x}=0$) or TM (transverse magnetic, i.e. $E_{y}=H_{z}=H_{x}=0$) modes. In 3D rectangular WGs, $E_{z}$ and $H_{z}$ do not vanish and the TE and TM modes give rise to hybrid $\mathrm{\vec{HE}}_{mp}$ and $\mathrm{\vec{EH}}_{mp}$ WG modes where, as already mentioned, $m$ and $p$ denote the number of nodes along the $\hat{x}$ and $\hat{y}$ directions, respectively. In real 3D rectangular WGs, however, the refractive index difference between the core and the cladding layer is usually small. As a consequence, the angle between the $\hat{z}$ axis and the propagation direction along the zig-zag path of the waves is also small~\cite{ebeling1993integrated}. This means that the WG modes are strongly polarized along the $\hat{x}$ or $\hat{y}$ direction and behave as quasi-TE or quasi-TM modes. This allows us to classify them according to the direction of the dominant transverse electric field component~\cite{ebeling1993integrated}: 
\begin{eqnarray}
\mathrm{HE~modes:} 
  \left\{
    \begin{array}{l}
      \left|E_{y}\right|\gg\left|E_{x}\right|, \left|E_{z}\right|\\
      \left|H_{x}\right|\gg\left|H_{y}\right|, \left|H_{z}\right|
    \end{array}
  \right.
\end{eqnarray}
and
\begin{eqnarray}
\mathrm{EH~modes:} 
  \left\{
    \begin{array}{l}
      \left|E_{x}\right|\gg\left|E_{y}\right|, \left|E_{z}\right|\\
      \left|H_{y}\right|\gg\left|H_{x}\right|, \left|H_{z}\right|.
    \end{array}
  \right.
\end{eqnarray}

The exact calculation of the modal field distributions $\vec{E}_{\nu}=\vec{E}_{mp}$ and $\vec{H}_{\nu}=\vec{H}_{mp}$ as well as the propagation constants $\beta_{\nu}$ is not possible for $\mathrm{\vec{HE}}_{mp}$ and $\mathrm{\vec{EH}}_{mp}$ WG modes, and only approximate numerical solutions can be found. Figure~\ref{fig: WGs_composition}(c) shows the simulated optical field distribution corresponding to the fundamental $\mathrm{HE_{00}}$ mode of an (Al,Ga)As rib WG structure assuming an optical wavelength $\lambda=910$~nm, $h=150$~nm, and $w=900$~nm. The field distribution was calculated numerically using the beam propagation method~\cite{1017609} in combination with the effective index method~\cite{ebeling1993integrated}. The guiding layer consists of a GaAs layer of thickness $H=300$~nm deposited on top of a 1500-nm-thick $\mathrm{Al_{0.2}Ga_{0.8}As}$ layer. The upper part of the guiding layer is surrounded by air ($n_{c}\approx 1$). The refractive indices considered in the calculation are 3.578 and 3.415 for the GaAs and $\mathrm{Al_{0.2}Ga_{0.8}As}$ layers, respectively. This results in an effective refractive index $n_{\mathrm{eff}}=3.45734$ for the fundamental $\mathrm{HE_{00}}$ mode at $\lambda=910$~nm. Throughout the remainder of this review, we will simply refer to $\mathrm{\vec{HE}}_{mp}$ and $\mathrm{\vec{EH}}_{mp}$ WG modes as TE or TM modes, respectively.

\section{Surface acoustic waves in piezoelectric semiconductors}\label{Surface acoustic waves in piezoelectric semiconductors}

\subsection{Surface acoustic modes on semiconductor structures}\label{Surface acoustic modes on semiconductor structures}

The different types of acoustic waves propagating along a homogeneous solid medium can be obtained after solving Newton's equation of motion for the particle displacement $\vec{u}$~\cite{oliner1994acoustic,Ash1985,1449830,royer2000elastic}: 
\begin{equation}
\nabla\cdot\boldsymbol{T}=\rho\frac{\partial^{2}\vec{u}}{\partial t^{2}},
\label{Elastic_wave_eq}  
\end{equation}
by taking into account the piezoelectric constitutive relations given by: 
\begin{align}
\begin{split}\label{Piezo_constitutive_relations}
\boldsymbol{T}&=\boldsymbol{c}\boldsymbol{S}-\boldsymbol{\sigma}\vec{F} \\
\boldsymbol{D}&=\boldsymbol{\sigma}\boldsymbol{S}+\boldsymbol{\epsilon}\vec{F}.
\end{split}
\end{align}

Here, $\boldsymbol{T}$ and $\boldsymbol{S}$ are the stress and strain tensors, whereas $\vec{F}$ and $\vec{D}$ are the piezoelectric and electrical displacements fields, respectively. In this paper, tensors will be denoted by bold characters, whereas vector fields will be denoted by an arrow. In addition, we will use a reference coordinate system defined by the unit vectors $\hat{x}$, $\hat{y}$, and $\hat{z}$ with $\hat{x}$ parallel to the wave propagation direction. In the case of surface modes, we will assume that $\hat{y}$ is perpendicular to the surface. 

As the speed of sound is small compared to the speed of light, the piezoelectric field $\vec{F}$ can be written in terms of the piezoelectric potential $\Phi_\mathrm{SAW}$ as $\vec{F}=-\vec{\nabla}\Phi_\mathrm{SAW}$ (quasi-static approximation). The material parameters involved in~\ref{Elastic_wave_eq} and \ref{Piezo_constitutive_relations} are the density ($\rho$), the elastic stiffness ($\boldsymbol{c}$), the piezoelectric ($\boldsymbol{\sigma}$), and the dielectric ($\boldsymbol{\epsilon}$) tensors, respectively. From these two equations, it can be inferred  that the mechanical wave is accompanied by an electrical field in piezoelectric materials , as both equations are coupled by the piezoelectric tensor $\boldsymbol{\sigma}$, with $\boldsymbol{c}$ and $\boldsymbol{\epsilon}$ accounting for the elastic and electrical properties of the material, respectively.

The solutions to~\ref{Elastic_wave_eq} and \ref{Piezo_constitutive_relations} correspond to plane elastic waves propagating in a piezoelectric medium with increased elastic constants $\boldsymbol{c'}=\boldsymbol{c}(1+\boldsymbol{K}^{2})$. Here, $\boldsymbol{K}^{2}=\boldsymbol{\sigma}^{2}/\boldsymbol{c\epsilon}$ is the electromechanical coupling constant tensor, which determines the electro-acoustic conversion efficiency in a piezoelectric material. This effect, which is usually referred as piezoelectric stiffening, increases the propagation velocity of plane acoustic waves due to piezoelectricity.  From the definition of $\boldsymbol{K}^{2}$ given above, it can be clearly seen that the phase velocity of the elastic waves depends simultaneously on the dielectric ($\boldsymbol{\epsilon}$), piezoelectric ($\boldsymbol{\sigma}$), and the elastic properties ($\boldsymbol{c}$) of the medium, which are, at the same time, strongly dependent on both the crystal cut and the propagation direction of the waves. In general, there are three independent bulk elastic modes for a given wave vector $\vec{k}$, with phase velocity $v_{i}=\sqrt{c'_{i}/\rho}$ (for $i=1,2,3$), where $c'_{i}$ are the increased elastic constants of the material in the propagation direction of the mode $i$. These waves are usually referred as bulk acoustic waves (BAWs). One of the resulting waves has its polarization vector (which indicates the direction of the particles displacement) approximately parallel to the direction of propagation defined by $\vec{k}$, and thus it is called quasi-longitudinal acoustic wave. The other two waves have polarization vectors almost perpendicular to the direction of propagation and are called quasi-transverse or quasi-shear acoustic waves. For isotropic materials as well as for propagation along high symmetry directions in a crystal, the polarization of the waves can be exactly transverse or parallel to the direction of propagation. In this case, the three waves become pure modes, which can be classified according to their particle displacement pattern as longitudinal (LA) or transverse (TA) acoustic modes.

In the presence of a free piezoelectric surface located at $x=x_{s}$, surface elastic modes localized near the surface can appear if the appropriated boundary conditions are satisfied. These modes propagate parallel to the interface with a phase velocity $v_\mathrm{SAW}$ determined by the crystal cut and propagation direction, and with displacement and potential amplitudes decaying rapidly towards the bulk. The continuity of the stress and electrical displacement field imposes the following boundary conditions at the surface: 
\begin{align}
\begin{split}\label{Boundary_conditions_homogeneous} 
&T_{jx}\left(x=x_{s}^{-}\right)=0 \\
&D_{x}\left(x=x_{s}^{-}\right)=D_{x}\left(x=x_{s}^{+}\right), 
\end{split} 
\end{align}
for $j=x,y,z$, where the symbols $(-)$ and $(+)$ make reference to the regions just below and above the surface defined by $x=x_{s}$, respectively. Surface acoustic modes correspond to solutions of~\ref{Elastic_wave_eq} and \ref{Piezo_constitutive_relations} which are confined near the surface (i.e., with amplitude decaying to zero as $\left|y\right|\rightarrow\infty$) and, simultaneously, fulfill the boundary conditions given by~\ref{Boundary_conditions_homogeneous}. These SAW modes have a linear energy \emph{versus} wave vector $(\vec{k}_\mathrm{SAW})$ dispersion relation. If the medium is not homogeneous but comprises two or more layers, the boundary conditions given by~\ref{Boundary_conditions_homogeneous} must be applied to the interfaces between the constituent layers. For two adjacent layers grown along $\hat{x}$ with interface at the plane $x=x_{int}$, the continuity of the stress and electrical displacement field imposes the following boundary conditions at the interface between the layers:
\begin{align}
\begin{split}\label{Boundary_conditions}
T_{jx}\left(x=x_{int}^{-}\right)=T_{jx}\left(x=x_{int}^{+}\right) \\
D_{x}\left(x=x_{int}^{-}\right)=D_{x}\left(x=x_{int}^{+}\right), 
\end{split}
\end{align}
for $j=x,y,z$, where the symbols $(-)$ and $(+)$ again make reference to the regions just below and above the interface defined by $x=x_{int}$, respectively. In general, the presence of a free surface at $x=x_{s}$ decreases the stiffness of the solid, causing $v_\mathrm{SAW}$ to be slightly lower than the phase velocity expected for BAWs. Moreover, because the amplitude of surface acoustic modes reduces with depth along $\hat{y}$, the electromechanical coupling constant tensor $\boldsymbol{K}^{2}$ must be replaced in this case by an effective counterpart $K_\mathrm{eff}^{2}$. Given the crystal cut and propagation directions, it can be approximated by $K_\mathrm{eff}^{2}\approx \left(v_\mathrm{SAW}-v_{0}\right)/v_\mathrm{SAW}$, where $v_{0}$ is the SAW phase velocity when the surface is shorted by a massless thin conducting metal film~\cite{PhysRevB.40.7874,1449830}. There are several types of surface acoustic modes, which have been thoroughly described in the literature~\cite{1449830,oliner1994acoustic,royer2000elastic}. Of particular interest are Rayleigh-type surface acoustic modes, which are a generalization of the well-known Rayleigh waves in isotropic media. 

Rayleigh-type SAWs result from the coupling between a longitudinal acoustic mode and a transverse acoustic mode polarized perpendicular to the surface (i.e., the acoustic modes are phase-shifted by $\pi/2$ with respect to each other). The particle displacement follows an elliptical path with the major axis  perpendicular to the surface plane defined by $x=x_{s}$. As expected for surface acoustic modes, the amplitude of the particle displacement decays exponentially with depth. For certain crystal cuts and propagation directions, the acoustic modes travel accompanied by a piezoelectric component that propagates perpendicular to the surface plane and therefore, can be excited electrically by means of IDTs, which will be discussed in Section~\ref{Excitation of surface acoustic waves}. As an example, the Rayleigh-type surface acoustic mode propagating along the $\left[110\right]$ direction of a $\left(001\right)$ GaAs surface is piezoelectric and can be excited electrically using IDTs. This makes these crystal cut and propagation directions particularly attractive for planar integrated acousto-optic devices on GaAs. On the contrary, the Rayleigh-type surface acoustic mode propagating along the $\left[100\right]$ direction of the same GaAs surface is non-piezoelectric. These non-piezoelectric modes can be excited by placing the IDTs on a piezoelectric film deposited on the substrate.

\subsection{Excitation of surface acoustic waves}\label{Excitation of surface acoustic waves} 

Surface acoustic modes are commonly excited using short laser pulses or electrically through the inverse piezoelectric effect. In the former case,~\cite{doi:10.1063/1.350747} a pulsed laser heats the target material, generating a local strain field (thermo-elastic effect) that propagates along the surface of the sample in the form of a SAW. In the latter~\cite{doi:10.1063/1.1754276,1449830,campbell1989surface,royer2000elasticII,hashimoto2000surface}, the surface modes are excited electrically using IDTs deposited on the surface of a piezoelectric material. Conversely, an IDT can also be used to detect an incident SAW beam by means of the direct piezoelectric effect. This twofold mode of operation, together with the planar format and the ease of fabrication using standard techniques such as optical lithography or electron-beam lithography, makes IDTs ideal for integrated acousto-optic functionalities.

An IDT comprises a planar periodic array of thin-film metal stripes or electrode fingers connected to two contact pads. Its resonance frequency, as well as the SAW wavelength ($\lambda_\mathrm{SAW}$), are defined by the spatial period of the electrode finger array. When a sinusoidal radio-frequency (RF) signal is applied between the two contact pads, the polarity of the electric field created between the electrodes alternatively changes from one electrode to the next. This changing electric field generates a periodic strain by means of the inverse piezoelectric effect, in such a way that each period of electrode fingers acts as a source of elastic waves. If the frequency of the applied RF signal equals the resonance frequency of the IDT, the propagating elastic waves generated at each period of the IDT add constructively and a powerful SAW beam can be generated. In the most common types of IDT, which are symmetric with respect to the middle plane, SAW beams are radiated both towards the left and right directions with equal amplitude and beam width determined by the overlap between adjacent electrode fingers. If, on the contrary, the frequency of the RF signal is shifted far from the resonance frequency, the interference is no longer completely constructive and the acoustic power carried by the radiated SAW beam is reduced. The grating structure of the array of electrode fingers may in principle also enable the excitation of BAWs towards the substrate. This is an undesirable effect, since the total power that is effectively converted into SAWs is considerably reduced. 

%%%%%%%%%%%%%%%%%%%%%%%%%%%%%%%%%%%%%%%%%%%%%%%%%%%%%%%%%%%%%%%%%%%%%%%%%%%%%%%%%%%%%%
\begin{figure*}[t!]
\begin{center}
		\includegraphics[width=0.95\textwidth]{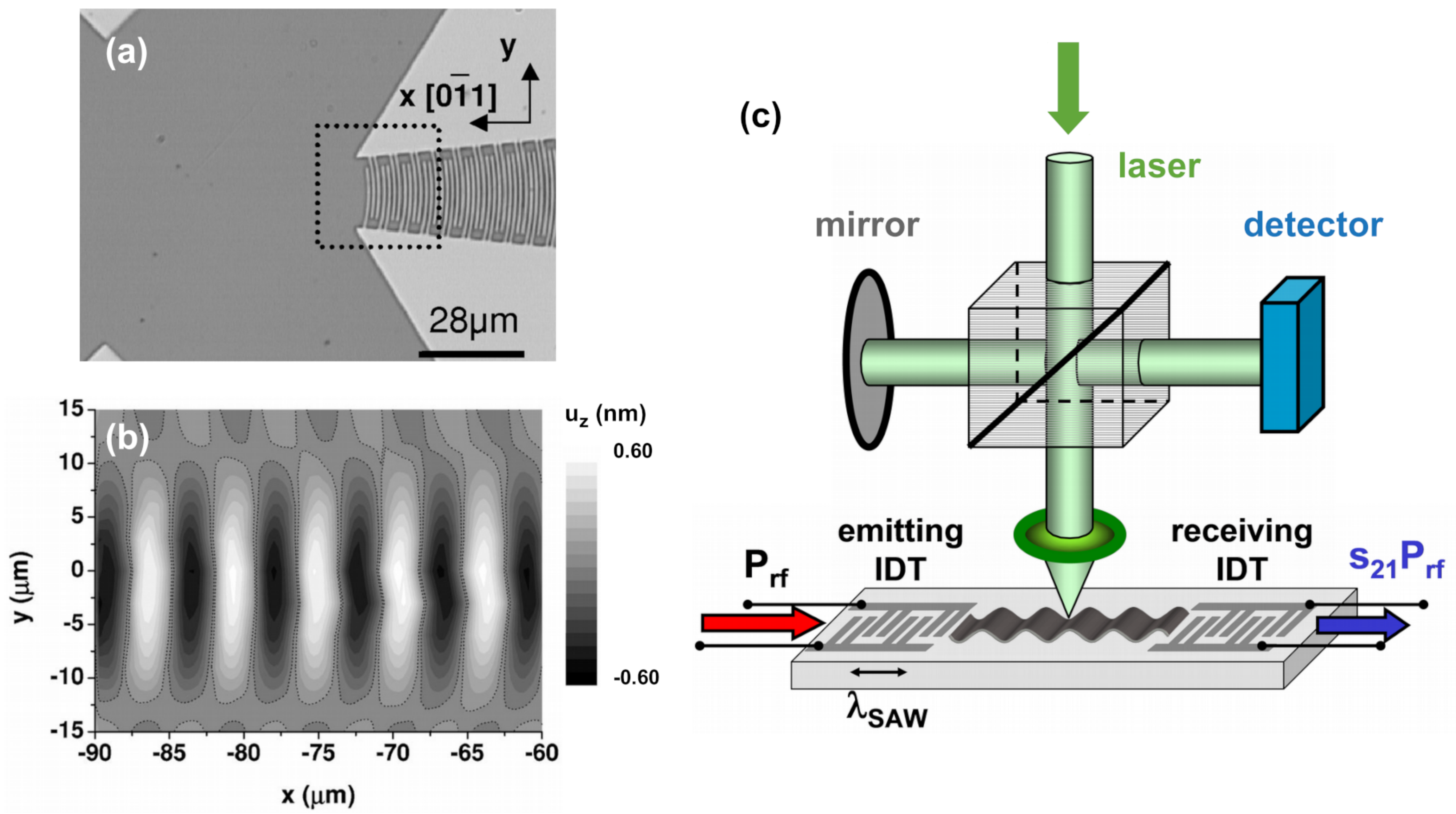}
	\caption{(a) Optical micrograph of the $18~\mu$m-wide aperture output of a double-finger focusing IDT. The dotted rectangle indicates the area scanned in the interferometry measurement shown in (b). This IDT was designed for an operating wavelength of $\lambda_{SAW}=5.6~\mu$m and $f_{SAW}=510$~MHz. The curved fingers are designed to follow the SAW group velocity curves. Adapted from~\cite{doi:10.1063/1.1625419}. (b) Vertical displacement $\left(u_{z}\right)$ measured by interferometry in the dotted area indicated in (a), excited with RF power $P_{RF}=10$~mW. Reproduced from~\cite{doi:10.1063/1.1625419}. (c) Sketch of the interferometry setup used to record the amplitude of $u_{z}$ induced by the SAW as shown in (b). The sample comprises a delay line with two IDTs for SAW generation (left side) and detection (right side), respectively.} %Adapted from~\cite{de2005modulation}.}
	\label{fig:Combination_focusing_IDT_de_Lima}
\end{center}
\end{figure*}
%%%%%%%%%%%%%%%%%%%%%%%%%%%%%%%%%%%%%%%%%%%%%%%%%%%%%%%%%%%%%%%%%%%%%%%%%%%%%%%%%%%%%%

The frequency window around the resonance frequency in which an electric signal can be efficiently converted into a SAW beam (i.e., bandwidth) is inversely proportional to the total length of the IDT and, thus, to the number of electrode fingers. In materials with weak piezoelectricity, such as GaAs, the SAW power radiated by each period of electrode fingers is usually small. In principle, this issue can be overcome by means of IDTs with a larger number of periods. However, the use of larger IDTs imposes severe technological limitations for very high operation frequencies (i.e., $>1$~GHz), as SAWs are back-reflected and scattered into bulk acoustic modes as they propagate through the IDT. Reflection losses increase dramatically with frequency, thus leading to an important reduction of the acoustic power that is converted into SAWs and limiting the length of the array of electrode fingers that can be effectively used in realistic high-frequency IDTs. 

The aforementioned reflection losses are particularly important in single-finger IDTs. This configuration, which is broadly used due to its simplicity, comprises two electrode fingers per period of width and spacing $\lambda_\mathrm{SAW}/4$. In this layout, the array of electrode fingers acts as a grating with periodicity $\lambda_\mathrm{SAW}/2$, fulfilling the condition for Bragg reflection. These reflections, although usually small, interfere constructively at the resonance frequency. As a result, increasing the number of finger pairs beyond a certain limit will not increase the irradiated acoustic power and, thus,  worsen the overall efficiency of the IDT. The standard technique employed to minimize reflection losses consists of using a double-electrode (sometimes also called split-finger) configuration. This configuration comprises four electrode fingers of width and spacing $\lambda_\mathrm{SAW}/8$ per period. In this case, the acoustic reflections produced between the electrode fingers during SAW propagation interfere constructively at a frequency different from the resonance frequency of the IDT, which results in an improved response with frequency. Narrower electrode fingers demand, however,  a higher lithographic resolution of the fabrication process. The suppression of Bragg reflection in the split-finger configuration enables higher precision and control over the resonance frequency of the IDT. In other cases, however, reflections at the electrode fingers can be exploited to design unidirectional IDTs in which the SAW beam is preferentially radiated along a single direction~\cite{1449830,campbell1989surface,hashimoto2000surface}. Some approaches for unidirectional IDTs reported so far make use of reflection grating adequately inserted in between the electrode finger array or floating electrode fingers. With this layout, the points of transduction and the sources of reflection are separated within each period and the interaction of the produced elastic waves with back-reflections allows for the unidirectional emission of SAWs. Unidirectional IDTs have been traditionally employed for low-loss SAW filters~\cite{4249186,1539925,171324} and, more recently, in sensing applications~\cite{0964-1726-15-6-003,1347-4065-47-5S-4065} or even for quantum SAW experiments at very low temperatures~\cite{doi:10.1063/1.4975803}.    

For all IDT configurations, the finite size of the finger overlap and the IDT aperture leads to diffraction effects that make the radiated SAW beam undergo some angular dispersion during propagation. These effects are important when the width of the IDT is only a few times $\lambda_\mathrm{SAW}$ and, in piezoelectric materials, they are strongly determined by the SAW propagation direction. In practice, diffraction in IDTs can be reduced by increasing the finger overlap as well as the IDT aperture. 

For certain applications, such as the acoustic modulation of the electronic and optical properties of low-dimensional structures like quantum dots (QDs)~\cite{doi:10.1063/1.2976135} or nanowires~\cite{doi:10.1021/nl203461m,doi:10.1021/nl1042775}, it may be necessary to generate a strong SAW beam focused on a small region of the sample. A focused SAW beam can be accomplished by properly shaping the electrode fingers of the IDT. In the simplest case of acoustically isotropic materials, focusing can be achieved by shaping the electrode fingers in the form of segments of circumference. For piezoelectric materials, in which the propagation properties of SAWs are strongly anisotropic, the situation is more complex and the best results are obtained if the IDT finger electrodes follow  lines of constant group velocity~\cite{doi:10.1063/1.336130,doi:10.1063/1.336131,doi:10.1063/1.1625419,1509798} rather than segments of circumference. This condition ensures that the group velocity of the SAW beam generated at each period of the IDT is directed towards the point of focus, producing a narrow collimated SAW beam. Besides the proper shaping of electrode fingers, de Lima \emph{et al.}~\cite{doi:10.1063/1.1625419} determined that additional focusing of the SAW beam could be achieved on GaAs by using Al for metallizing the IDTs, this way achieving SAW beams with full width at half maximum (FWHM) less than $3\lambda_\mathrm{SAW}$. This is because, in this case, the acoustic velocity at the electrode finger array is lower than that at the contact pads and the SAW mode is confined within the former, which acts as a lateral acoustic WG. Figure~\ref{fig:Combination_focusing_IDT_de_Lima}(a) shows a top-view micrograph of a double-finger focusing IDT reported in~\cite{doi:10.1063/1.1625419}, designed for an operation wavelength of $\lambda_{SAW}=5.6~\mu$m and frequency $f_{SAW}=510$~MHz on a GaAs wafer. The dotted rectangle indicates the
area scanned in the interferometry measurement shown in (b). The IDT aperture is $18~\mu$m. Figure~\ref{fig:Combination_focusing_IDT_de_Lima}(b) shows the vertical displacement $\left(u_{z}\right)$ recorded by interferometry at the output of a double-finger focusing IDT designed for an operation wavelength of $\lambda_\mathrm{SAW}=5.6~\mu$m (corresponding to $f_\mathrm{SAW}=510$~MHz) on a GaAs wafer, for a fixed phase of the SAW and an excitation RF power $P_{RF}=10$~mW~\cite{doi:10.1063/1.1625419}. A sketch of the employed interferometry setup, including the sample which comprises a delay line with two IDTs for SAW generation (left side) and detection (right side), is shown in Figure~\ref{fig:Combination_focusing_IDT_de_Lima}(c).  

Finally, one should also mention that there are alternative approaches for SAW generation such as the use of a periodically-poled ferroelectric material has been demonstrated on $\mathrm{LiNbO_{3}}$ by Yudistira \emph{et al.}~\cite{doi:10.1063/1.3190518,doi:10.1063/1.3599569,Yudistira:10,6566094}.

\section{Acousto-optical interaction in semiconductor structures}\label{Acousto-optical interaction in semiconductor structures}

\subsection{Effects of strain and piezoelectric fields on the band edges}\label{Effects of strain and piezoelectric fields on the band edges}

\begin{figure*}[h!]
\centering
\includegraphics[width=0.96\textwidth]{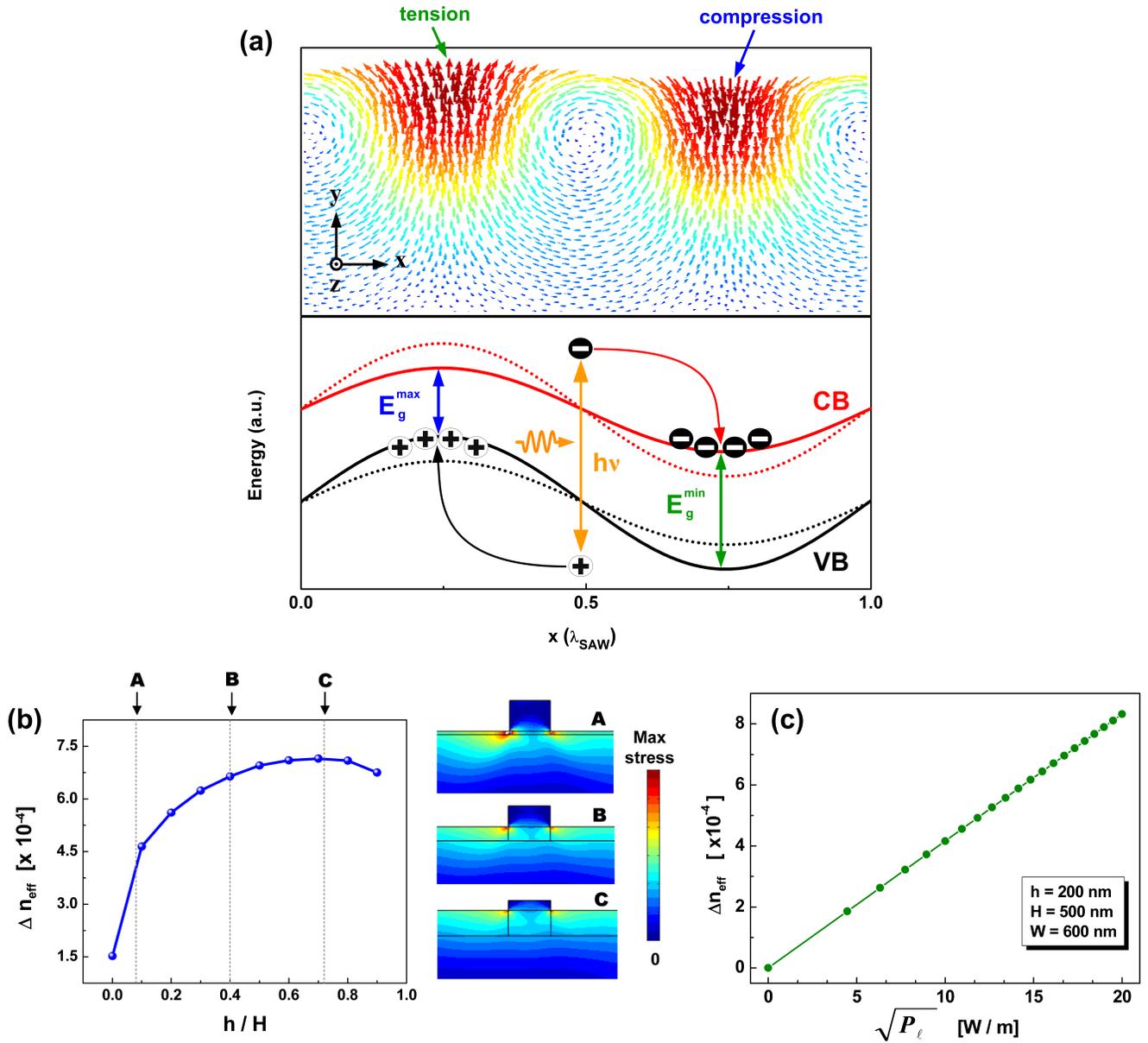}
\caption{(a) Upper panel: instantaneous displacements of the elementary volumes of solid by a SAW which propagates in a semiconductor crystal. Lower panel: modulation of the conduction band (CB) and the valence band (VB) edges of the crystal by the SAW strain. The piezoelectric potential $\left(\Phi_\mathrm{SAW}\right)$ associated with the strain introduces a type-II modulation of the band edges. The regions of maximum compression and tension induced by the strain superimposes a small modulation of the band gap with maximum and minimum values given by $E_{g}^{max}$ and $E_{g}^{min}$, respectively (type-I modulation). The piezoelectric field $\left(\Phi_\mathrm{SAW}\right)$ induces a spatial separation of carriers generated e.g. by optical means (type-II modulation). (b) Calculated effective refractive index change $\Delta n_{\mathrm{eff}}$ of the fundamental TE mode of (Al,Ga)As WGs with different $h/H$ ratios assuming an acoustic power density $P_{\ell}=200$~W/m, and plots of $S_{xx}$ for $h/H\approx 0.1$ (A), $=0.4$ (B), and $\approx 0.7$ (C). (c) Calculated effective refractive index change $\Delta n_{\mathrm{eff}}$ of the fundamental TE mode of an (Al,Ga)As WG for different linear acoustic power densities $P_{\ell}$ assuming $h=200$~nm, $H=500$~nm, and $W=600$~nm. (a) and (b) reproduced from~\cite{b950586b776740e292ffa6e3bb8b2a0e}.}
\label{fig:Modulation_band_gap}
\end{figure*}

As we have seen in Section~\ref{Surface acoustic modes on semiconductor structures}, SAWs are essentially mechanical waves with most of their power confined near the surface, which travel accompanied by an electrical field if the crystal is piezoelectric. In a semiconductor, the band structure under a SAW becomes modified through two different mechanisms as indicated in Figure~\ref{fig:Modulation_band_gap}. The first one is the deformation potential $\Xi_{ij}$ modulation of the band gap $\Delta E_{g}^{S}$ caused by the periodic local variations of the volume and symmetry of the crystal induced by the SAW strain field $S_{ij}$ (cf. Figure~\ref{fig:Modulation_band_gap}, upper panel)~\cite{yu2010fundamentals}:
\begin{equation}\label{Deformation potential}
\Delta E_{g}^{S}=E_{g}^{max}-E_{g}^{min}=2\Xi_{ij}S_{ij},
\end{equation}
where $E_{g}^{min}$ and $E_{g}^{max}$ correspond to the band gap values at the region of maximum tension and compression, respectively, created by the SAW field, as shown in Figure~\ref{fig:Modulation_band_gap}, lower panel. Here, the superscript $S$ has been placed to emphasize the strain field contribution to the band gap modulation through the elasto-optic effect (type-I modulation). The second mechanism is related to the piezoelectric potential $\Phi_\mathrm{SAW}$ created by the strain field in piezoelectric materials, which induces an indirect electro-optic modulation of the band structure (type-II modulation).

%%%%%%%%%%%%%%%%%%%%%%%%%%%%%%%%%%%%%%%%%%%%%%%%%%%%%%%%%%%%%%%%%%%%%%%%%%%%%%%%%%%%%%%%%%%%%%%%%%%%%%%%%%%%%%%%%%%%
%%%%%%%%%%%%%%%%%%%%%%%%%%%%%%%%%%%%%%%%%%%%%%%%%%%%%%%%%%%%%%%%%%%%%%%%%%%%%%%%%%%%%%%%%%%%%%%%%%%%%%%%%%%%%%%%%%%%

Defining the inverse of the dielectric tensor as $\boldsymbol{B}=\boldsymbol{\epsilon}^{-1}$ so that $\epsilon_{ij}B_{jk}=\delta_{ik}$, and considering only small changes $\Delta B_{ij}$ in its components (which is true for most practical applications), the variation $\Delta B_{ij}^{S}$ caused by the dynamic SAW strain relates linearly to the amplitude of the strain field according to the following expression~\cite{narasimhamurty2012photoelastic,royer2000elasticII}:
\begin{equation}\label{Elasto_optic_effect}
\Delta B_{ij}^{S}=p_{ijkl}S_{kl},
\end{equation}
where $p_{ijkl}$ are components of the fourth-rank elasto-optical tensor $\boldsymbol{p}$, which quantifies the strength of the interaction between photons and elastic waves. As discussed before, in a piezoelectric material, the SAW dynamic strain field also induces an indirect electro-optic modulation in the semiconductor band gap. The variation $\Delta B_{ij}^{E}$, where the superscript $E$ stands for electro-optic contribution, can be expressed in terms of the electric field $\vec{F}=-\vec{\nabla}\Phi_\mathrm{SAW}$ as~\cite{narasimhamurty2012photoelastic,royer2000elasticII}:
\begin{equation}\label{Electro_optic_effect}
\Delta B_{ij}^{E}=r_{ijl}F_{l},
\end{equation}
where $r_{ijl}$ are components of the third-rank electro-optical tensor $\boldsymbol{r}$. The latter quantifies the changes in the refractive index associated to the electric field $\vec{F}$ generated by the strain field. Therefore, in the most general case, the total variation $\Delta\boldsymbol{B}$ can be explicitly expressed as the sum of the elasto-optic and the indirect electro-optic contributions:
\begin{equation}\label{Sum_effects}
\Delta\boldsymbol{B}=\Delta\boldsymbol{B}^{S}+\Delta\boldsymbol{B}^{E}
\end{equation}

Both modulation mechanisms are relevant for acousto-optical modulation. However, for optical wavelengths away from electronic resonances, the elasto-optical modulation of the dielectric tensor $\boldsymbol{\epsilon}$ associated with the deformation potential $\Xi_{ij}$ mechanism usually dominates over the indirect electro-optical contribution related with the piezoelectric potential $\Phi_\mathrm{SAW}$. The electro-optical contribution may, however, become important in the presence of excitons or free carriers induced by optical fields or doping (see for instance~\cite{PhysRevLett.87.136403,PhysRevB.66.165330,PhysRevLett.94.226406}).

\subsection{Acousto-optical modulation of 3D optical waveguides}\label{Acousto-optical modulation of 3D optical waveguides}

One of the most traditional applications in acousto-optics consists of using SAWs to diffract an incoming light beam in a Bragg cell. In these devices, the propagating dielectric grating generated by the dynamic SAW strain field is used to scatter a light beam which propagates transversely to the acoustic beam in a planar 2D slab optical WG, enabling important applications such as modulators, frequency shifters, or spectrum analyzers, among others~\cite{korpel1996acousto,tsai2013guided}. In devices built upon 3D rectangular WGs, the SAW propagates perpendicularly to the WG and modulates its physical dimensions and its refractive index distribution. This results in a change in the effective refractive index $n_{\mathrm{eff},\nu}$ as given by~\ref{Effective index} that is sensed by WG mode $\nu$. 

In acousto-optical devices, the physical dimensions of the WGs are normally chosen so that only a single guided WG mode is supported. This allows us to disregard the interaction between SAWs and higher order (i.e. $\nu\geq 1$) optical WG modes, which considerably simplifies the modeling of the devices~\cite{b950586b776740e292ffa6e3bb8b2a0e}. This has also the effect of considerably enhancing the SAWs and the optical fields, as shown by D\"{u}hring and Sigmund~\cite{doi:10.1063/1.3114552}. Moreover, the WG width is usually chosen to be much smaller than $\lambda_{SAW}$ in such a way that the SAW-induced refractive index modulation can be considered constant across the WG cross section. This is important for devices whose operation principle relies upon fixed acoustic phases between the modulated WGs, as is the case with interferometer devices, which will be discussed in Section~\ref{Interferometric acousto-optic devices}.  

The SAW-induced phase shift $\Delta\Phi$ undergone by the light beam propagating in an optical WG supporting a single guided mode can be expressed as~\cite{doi:10.1063/1.2354411}:
\begin{equation}\label{Optical_phase}
\Delta\Phi=k_{0}\ell_{int}\Delta n_\mathrm{eff,0}=a_{p}\sqrt{P_{IDT}},
\end{equation}
where $k_{0}=2\pi/\lambda_{}$ and $\lambda_{}$ denote the light wave vector and wavelength in vacuum, respectively, $\Delta n_\mathrm{eff,0}\equiv\Delta n_\mathrm{eff}$ is the effective refractive index change of the fundamental WG mode due to the acousto-optical interaction, and $\ell_{int}$ is the interaction length between the SAW and the optical fields. The second equality in~\ref{Optical_phase} states that the total SAW-induced phase shift $\Delta\Phi$ is proportional to the square root of the nominal RF power applied to the IDT $P_{IDT}$. The proportionality constant $a_{p}$ is dependent on the materials elasto-optical properties as well as on the overlap between the optical and acoustic fields in the modulated region.

The dielectric modulation induced by the dynamic SAW strain (cf.~\ref{Elasto_optic_effect}) induces small changes $\Delta n_\mathrm{eff}$ in the effective refractive index of the fundamental WG mode $n_\mathrm{eff,0}\equiv n_\mathrm{eff}$, with typical values of $\Delta n_\mathrm{eff}/n_\mathrm{eff}\approx10^{-4}-10^{-3}$. As a consequence, in order to accomplish sizeable optical phase shifts, it is necessary to optimize the acousto-optical interaction. The most straightforward way consists of increasing the interaction length $\ell_{int}$ between the SAWs and the guided optical fields, with typical values of several $\lambda_\mathrm{SAW}$. Moreover, the geometry of the WG strongly determines the overlap between the optical and acoustic fields in the modulated region and is, therefore, also a determining factor in the efficiency of the acousto-optical modulation as demonstrated by Barretto in~\cite{b950586b776740e292ffa6e3bb8b2a0e}. Figure~\ref{fig:Modulation_band_gap}(b) shows the calculated effective refractive index change $\Delta n_{\mathrm{eff}}$ of the fundamental TE mode of (Al,Ga)As WGs with different $h/H$ ratios assuming a linear acoustic power density $P_{\ell}=200$~W/m, showing that $\Delta n_{\mathrm{eff}}$ is maximized for $h/H\approx 0.7$. This behaviour can be better understood by looking at the plots of the horizontal strain $S_{xx}$ calculated for $h/H\approx 0.1$ (A), $=0.4$ (B), and $\approx 0.7$ (C). Here, it can be seen that the strain is maximum at the lower corners of the WG, where there is not much guided optical power power [cf.~Figure~\ref{fig: WGs_composition}(c)]. This effect is particularly important in deeply-etched WGs with low $h/H$ such as the one depicted in (A). This causes that an important fraction of the total SAW power does not interact with the optical fields, reducing therefore the efficiency of the acousto-optic interaction and explaining the lower values of $\Delta n_{\mathrm{eff}}$ in the left plot. Besides optimizing the WG geometry, these limitations can also be overcome by burying the WGs close to the surface, making it possible to increase up to ten times the efficiency of the acousto-optical modulation~\cite{doi:10.1063/1.3114552}. Figure~\ref{fig:Modulation_band_gap}(c) shows the effective refractive index change $\Delta n_{\mathrm{eff}}$ as a function of the linear acoustic power densities $P_{\ell}$, calculated for the fundamental TE mode of the WG structure labeled (B) in Figure~\ref{fig:Modulation_band_gap}(b), assuming $h=200$~nm, $H=500$~nm, and $W=600$~nm. Here, the linear dependence of $\Delta n_{\mathrm{eff}}$ with the SAW power is clearly visible, as can be expected from the acousto-optical origin of the interaction (cf.~\ref{Optical_phase}).

\section{Ring resonators modulated by surface acoustic waves}\label{Ring resonators modulated by surface acoustic waves}

Monolithic ring resonators are interesting building blocks for integrated photonic circuits. Some of their applications include optical switching~\cite{974166,1159054,Wen:11}, integrated bio-sensors~\cite{Ksendzov:05,5232865}, or laser resonators~\cite{doi:10.1063/1.1420585,998697}, among others. In their basic configuration, they comprise two straight access WGs (also known as bus WGs) which are coupled to a ring WG through the evanescent optical field. When an optical signal comprising $N$ wavelength components $\lambda_{i}$ (with $i=1,...,N$) is launched in one of the bus WGs, only the wavelength satisfying the resonant condition will couple to the ring WG~\cite{rabus2007}: 
\begin{equation}
n_\mathrm{eff}^{R}L_{R}=m\lambda_{i}, 
\label{Ring_resonator_resonance_condition} 
\end{equation}
where $n_\mathrm{eff}^{R}$ and $L_{R}$ are the effective refractive index and the perimeter of the ring WG, respectively, and $m$ is an integer. Assuming, for instance, that two different optical wavelengths $\lambda_{1}$ and $\lambda_{2}$ are launched into bus WG 1 and that only $\lambda_{1}$ fulfills the resonance condition given by~\ref{Ring_resonator_resonance_condition}, only the latter will be temporarily stored in the ring WG and coupled to bus WG 2. Accordingly, wavelength $\lambda_{2}$ will be fully transmitted for detection at the opposite end of bus WG 1. 

This passive response can be dynamically tuned by use of piezoelectrically-generated SAWs~\cite{FAN201462,fan2016process}. In this case, the SAW strain modifies the transmission spectrum of the devices by inducing changes in $n_\mathrm{eff}^{R}$ as given by~\ref{Optical_phase}. Moreover, if the SAW propagates perpendicularly to the region in which the bus WGs couple to the ring WG, the coupling distance between the latter becomes also modulated with the SAW period $T_\mathrm{SAW}$ leading to additional tuning of the transmission~\cite{fan2016process}.

%%%%%%%%%%%%%%%%%%%%%%%%%%%%%%%%%%%%%%%%%%%%%%%%%%%%%%%%%%%%%%%%%%%%%%%%%%%%%%%%%%%%%%%%%%%%%%%%%%%%%%%%%%%%%%%%
\begin{figure*}[t!]
\centering
\includegraphics[width=0.90\textwidth]{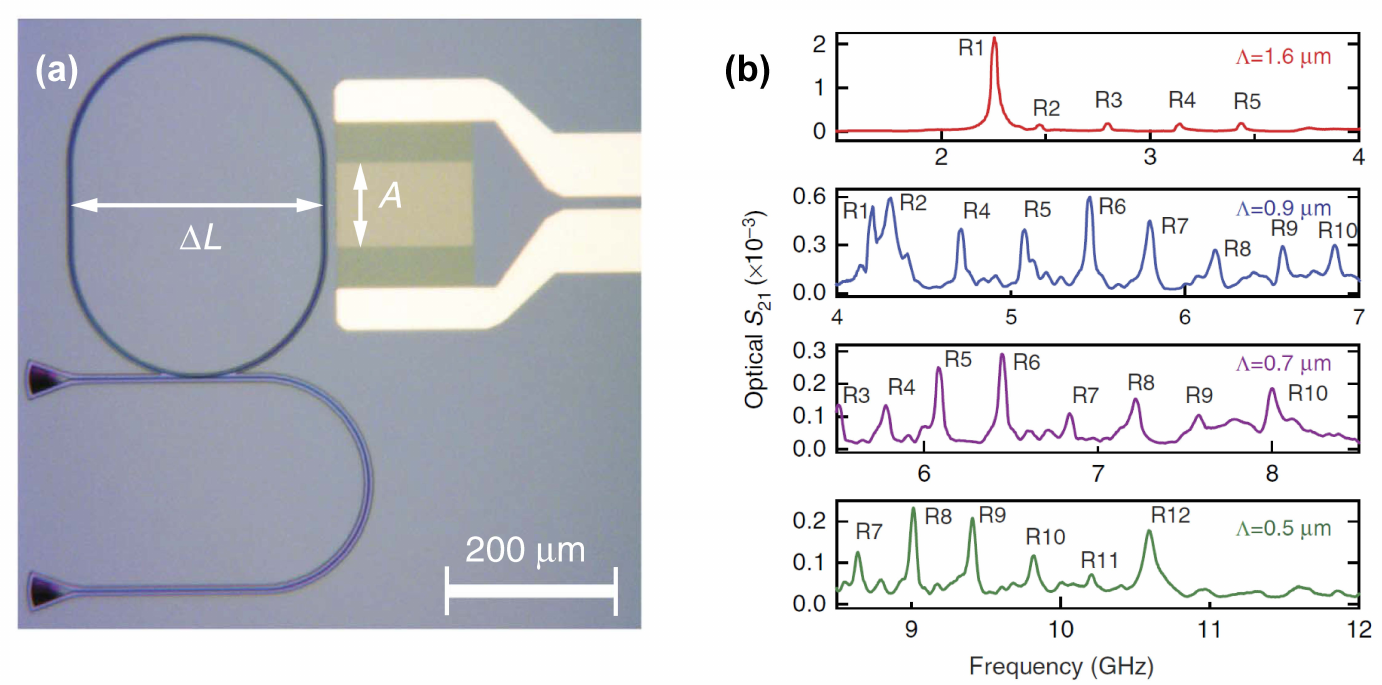}
\caption{(a) Top-view micrograph of a SAW-tuned ring resonator investigated by Tadesse and Li~\cite{tadesse2014sub,tadesse2016}. Here, $\Delta L$ is the separation between the modulated straight WG sections of the ring resonator and $A$ is the DIT aperture. (b) Experimental spectrum of the $S_{21}$ optical transmission coefficient of the same device, recorded for different $\lambda_\mathrm{SAW}$ ($\Lambda$ in the drawing). Here, $R_{n}$ denote the $n^{th}$ Rayleigh SAW mode. The maximum modulation frequency is 10.6~GHz, corresponding to $R_{12}$ with $\lambda_\mathrm{SAW}=0.5~\mu$m. Reproduced from~\cite{tadesse2014sub}.}
\label{fig:SAW_tuning_ring_resonators}
\end{figure*}
%%%%%%%%%%%%%%%%%%%%%%%%%%%%%%%%%%%%%%%%%%%%%%%%%%%%%%%%%%%%%%%%%%%%%%%%%%%%%%%%%%%%%%%%%%%%%%%%%%%%%%%%%%%%%%%%

An alternative interesting mechanism for further increasing the acousto-optical modulation is that of sub-optical wavelength modulation, by which the acoustic wavelength $\lambda_\mathrm{SAW}$ is reduced to much less than the optical wavelength $\lambda_{}$: $\lambda_\mathrm{SAW}\ll\lambda_{}$, as demonstrated by Tadesse and Li~\cite{tadesse2014sub}. In this regime, it is possible to achieve a large overlap between the acoustic and the optical modes, thus optimizing the acousto-optical interaction. The authors have demonstrated that the optimum overlap takes place when the SAW wavelength $\lambda_\mathrm{SAW}$ is equal to twice the lateral size of the optical mode. On the contrary, the acousto-optical modulation almost vanishes when $\lambda_\mathrm{SAW}$ equals the lateral size of the optical mode, as the acousto-optical contribution from the compressive and tensile strain components (cf.~Figure~\ref{fig:Modulation_band_gap}) cancel each other. This is a direct consequence of the number of SAW nodes and anti-nodes that fit within the WG width i.e., the resulting acousto-optical modulation is non-zero only if their number is different. As can be expected, the situation varies periodically for a fixed WG width as $\lambda_\mathrm{SAW}$ is reduced more and more. The devices investigated by Tadesse and Li in~\cite{tadesse2014sub,tadesse2016} were fabricated on a $\mathrm{AlN/SiO_{2}/Si}$ multilayer structure. A top-view micrograph of the devices is shown in Figure~\ref{fig:SAW_tuning_ring_resonators}(a). These comprise two access ports where the light is in/out coupled using grating couplers~\cite{doi:10.1063/1.1653091} which are subsequently coupled to a ring resonator. The latter consists of two straight WG sections separated by a distance $\Delta L$ and linked by two arc WGs. When the wavelength of the input optical field matches the resonance condition of the ring (cf.~\ref{Ring_resonator_resonance_condition}), the transmission displays a sharp minimum. The IDT, with an operation wavelength $\lambda_\mathrm{SAW}$ ranging from $0.4~\mathrm{to}~1.6~\mu$m depending on the case, is placed so that the generated SAWs propagate perpendicular to the straight WG sections. In order to maximize the acousto-optical interaction, the aperture $A$ of the IDT equals the length of the straight WG sections. The $\mathrm{AlN/SiO_{2}/Si}$ multilayer structure of the devices make each SAW mode propagate with a different acoustic velocity $v_\mathrm{SAW}$ along each layer, leading to strong dispersion effects. In this way, the authors demonstrated that higher-order SAW modes dominate in the limit of short acoustic wavelengths, in which $2\pi(d/\lambda_\mathrm{SAW})>1$ with $d$ the thickness of the AlN piezoelectric overlayer. Figure~\ref{fig:SAW_tuning_ring_resonators}(b) shows the recorded transmission coefficient $S_{21}$ for a WG width of 800~nm and IDTs with operation wavelengths $\lambda_\mathrm{SAW}=1.6,0.9,0.7,~\mathrm{and}~0.5~\mu$m ($\Lambda$ in the picture). Here, $R_{n}$ stands for the higher-order Rayleigh SAW modes, with $n$ the modal number. In the measured spectrum, the highest modulation frequency is 10.6~GHz, corresponding to $R_{12}$ with acoustic wavelength $\lambda_\mathrm{SAW}=0.5~\mu$m.

\section{Acousto-optic interferometer devices}\label{Interferometric acousto-optic devices}

The general basis of operation of the acousto-optical interferometer devices described in this section is as follows: a guided optical beam is first split into two or more guided beams which are delivered to the modulated region of the device. The latter comprises additional WG sections with their refractive indices modulated by one or more SAW beams propagating perpendicularly to them. The SAW-induced optical phase delays can be precisely tailored by choosing the appropriate spacing between the modulated WG sections, with respect to $\lambda_\mathrm{SAW}$. The phase-delayed optical fields are then combined, and the subsequent interference allows for important functionalities such as optical modulation, by which the data signal can be transferred to the laser-generated carrier signal in optical communications, or switching. Furthermore, by imposing wavelength dispersion at the modulated WGs of the devices, very compact reconfigurable wavelength-division multiplexing (WDM) devices can be obtained.

Although the explanation is focused on the acousto-optical modulation, the concepts introduced in this section can be readily applied to other modulation techniques such as thermo- or electro-optical devices, enabling compact high-speed reconfigurable data routers which can be cascaded, arranged in matrices, or used in combination with standard WDM devices.

\subsection{Acousto-optic Mach-Zehnder modulators}\label{Mach-Zehnder acousto-optic modulators and switches}

%%%%%%%%%%%%%%%%%%%%%%%%%%%%%%%%%%%%%%%%
\begin{figure*}[t!]
	\centering
		\includegraphics[width=0.90\textwidth]{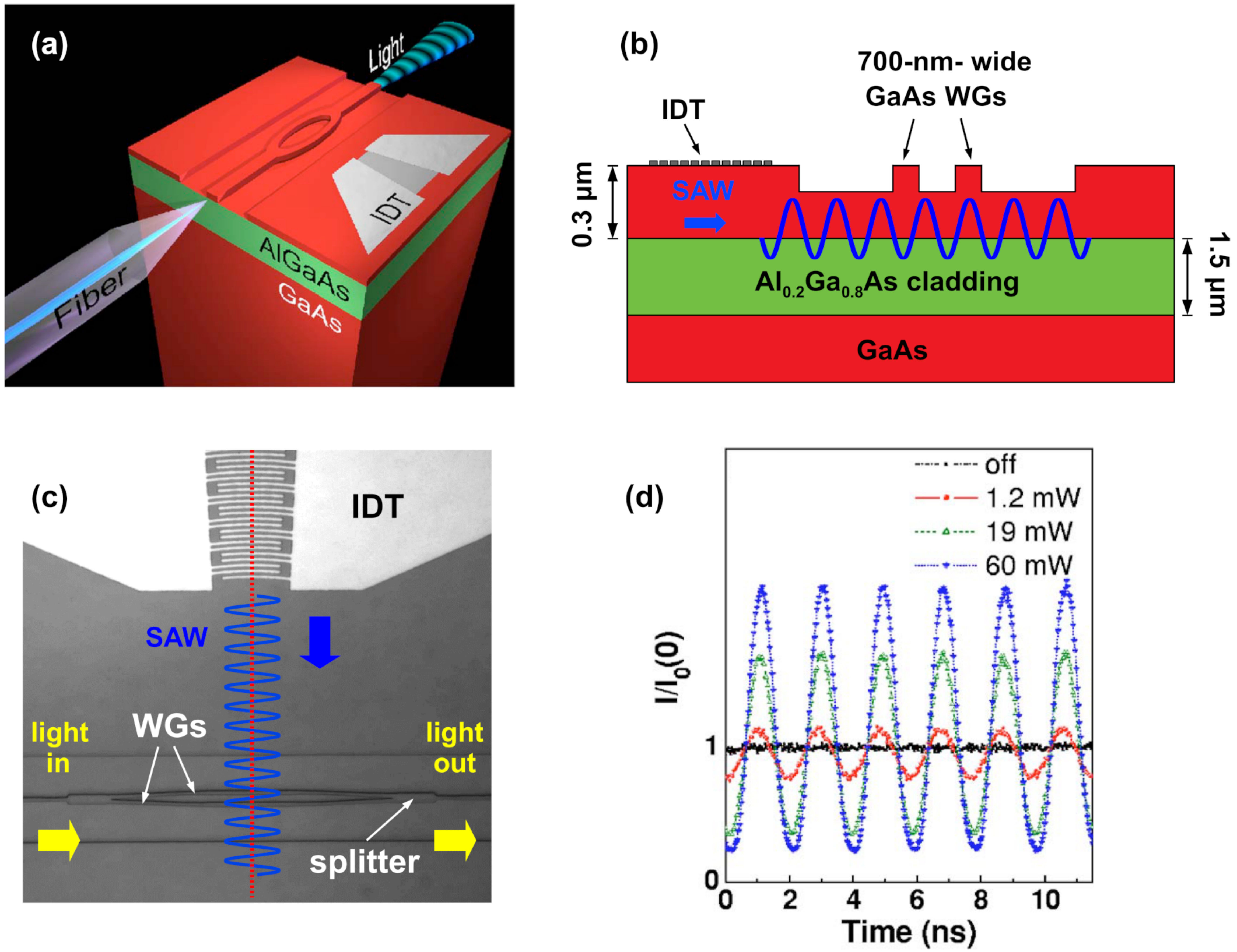}
	\caption{Acoustic Mach-Zehnder interferometer (MZI): (a) schematic configuration of a MZI modulated by SAWs generated by a focusing IDT as demonstrated by de Lima~\emph{et al.}~\cite{doi:10.1063/1.2354411}. Reproduced from~\cite{doi:10.1063/1.2354411}. (b) Corresponding layer cross-section of a device consisting of (Al, Ga)As layers grown by molecular beam epitaxy. (c) Top-view micrograph of an acoustic MZI using a focusing IDT designed for a wavelength of $5.6~\mu$m with an interaction length of $15~\mu$m. (d) Light transmission under different acoustic excitation power $P_IDT$ measured on the acoustic MZI in (a). The transmitivity is normalized to the transmission in the absence of acoustic excitation. Reproduced from~\cite{doi:10.1063/1.2354411}.}
	\label{aMZI}
\end{figure*}
%%%%%%%%%%%%%%%%%%%%%%%%%%%%%%%%%%%%%%%

The application of piezoelectrically-generated SAWs to control the response of a Mach-Zehnder interferometer (MZI) was first suggested by Gorecki~\emph{et al.}~\cite{Gorecki:97,doi:10.1117/12.281184,Bonnotte:99}. The device demonstrated by the authors consisted of a regular MZI structure comprising two symmetrical Y-junctions built upon rib WGs on a Si substrate. In order to generate a SAW beam, the IDT was placed on a thin piezoelectric ZnO overlayer with a high electromechanical coupling factor. In their layout, phase modulation was accomplished when the refractive index of one of the arms was modulated by a SAW beam propagating perpendicularly to the MZI structure. The other arm, which was used as the interferometer reference arm, was isolated from the SAW beam by an isolation trench.

A similar concept was employed by de Lima~\emph{et al.}~\cite{doi:10.1063/1.2354411} [cf.~Figure~\ref{aMZI}(a)]. Their work, however, introduced two main modifications to the work by Gorecki~\emph{et al.} in order to increase the modulation efficiency and reduce the interaction length. First, the modulation efficiency was doubled by modulating simultaneously both MZI arms with opposite phases using the same SAW beam. This can be clearly appreciated in Figure.~\ref{aMZI}(b), which displays the layer cross-section of the demonstrated MZI comprising (Al,Ga)As layers grown by molecular beam epitaxy on a GaAs $\left(100\right)$ wafer. The optical ridge WGs were structured by etching the 300~nm-thick GaAs layer grown on a 1500~nm-thick Al$_{0.2}$Ga$_{0.8}$As cladding layer. For that purpose, the two WG arms were displaced from each other by an uneven multiple of the $\lSAW/2$. In general, the effective refractive index in $j^{th}$ WG arm $n_\mathrm{eff}\left(j\right)$ can be written as a function of the SAW-induced refractive index change $\Delta n_\mathrm{eff}$ according to the following expression~\cite{Crespo-Poveda:16,doi:10.1117/12.2208868}: 
\begin{equation}
n_\mathrm{eff,j}\left(t\right)=n_\mathrm{eff}^{0}+\kappa_{j}\Delta n_\mathrm{eff}\cos\left(\omega_\mathrm{SAW}t\right),
\label{Neff_change_MZI} 
\end{equation}
where $n_\mathrm{eff}^{0}$ is the unperturbed effective refractive index, $\kappa_{j}$ are numerical factors that account for the amplitude and the phase of the SAW modulation in the $j^{th}$ arm, and $t$ is the time. As the WG arms are in this case modulated with the same SAW amplitude but opposite phase, we have that $\kappa_{1,2}=\pm1$ for the modulated WG 1 and 2, respectively. 

The second modification of the original design by Gorecki~\emph{et al.} aimed at reducing the acousto-optic interaction length in order to make very compact WG  structures. It consisted in using a focusing IDT to generate a narrow and intense SAW beam on the two arms of the MZI interferometer. Figure~\ref{aMZI}(c) displays an optical micrograph of the acoustic MZI showing the split-finger focusing IDTs designed for an acoustic wavelength $\lSAW=5.6~\mu$m (corresponding to a fundamental SAW frequency $\fSAW=520$~MHz). These IDTs generate a SAW beam with a  FWHM $\ell\approx15~\mu$m collimated along hundreds of $\mu$m. The figure also shows the two arms of the MZI WGs, which were displaced along the SAW propagation direction by $2.5\lambda_\mathrm{SAW}$ (i.e. by an odd multiple -5- of $\lambda_\mathrm{SAW}/2$). 

The modulation properties of the acoustic MZI were measured by coupling a 890~nm light beam to the WGs using optical fibers as illustrated in Figure~\ref{aMZI}(a). The IDT was excited with RF powers up to $P_{IDT}=60$~mW. Figure~\ref{aMZI}(c) displays time-resolved  transmission ($T$) traces recorded for different acoustic powers (the transmission values are normalized to the transmission recorded in the absence of a SAW). The transmission signal is modulated at the frequency of the SAW: this behavior is attributed to a fabrication-related phase difference in the MZI arms (for a perfectly symmetric MZI, the small signal modulation should be in the second harmonic of the SAW frequency). The modulation  amplitude reaches depths close to 100\%. 
 The time traces are slightly asymmetric due to the presence of higher harmonic components of the SAW frequency $f_\mathrm{SAW}$. By comparing the theoretical and the experimental results, the authors estimated the value of the proportionality constant (cf.~\ref{Optical_phase}) to be $a_{p}=7.3\times10^{-2}~\mathrm{rad/\sqrt{mW}}$.

\subsection{Acousto-optic multiple interference devices - AOMIDs}\label{Acousto-optic multiple interference devices - AOMIDs}

\begin{figure*}[t!]
	\centering
		\includegraphics[width=0.95\textwidth]{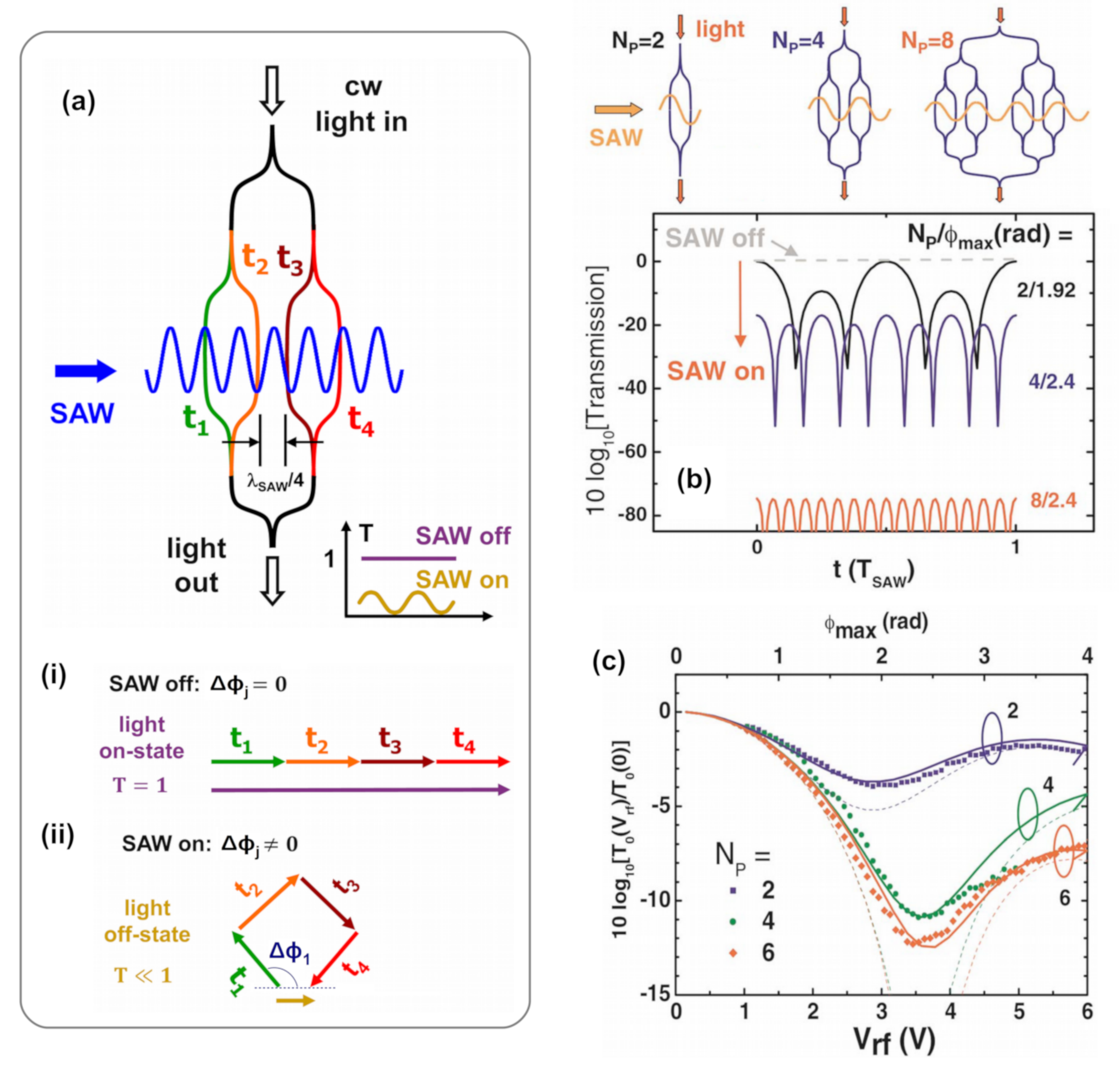}
	\caption{(a) Principle of operation on an acousto-optic multiple interference device (AOMID) with $N_p=4$  parallel WGs. A propagating SAW modulates the transmission $t_j$ ($j=1, 2, 3, 4$) of the light field in four parallel WGs separated by a multiple of the $\lSAW/N_p$. $t_j$ can be visualized as a vector with unity amplitude and phase $\Delta\Phi_j$ determined by the position of WG $j$.  In the absence of a SAW (i), the total transmission intensity $T$ given by~\ref{EqAOMID} is unit, leading to the light-on state of the switch. In the presence of a SAW (ii), the vectors $\vec{t}_j$ will add to yield a much lower transmission, thus blocking light propagation (light-off state). (b) Calculated time-dependent transmission for AOMIDs with different number of parallel channels $N_p$ determined for the phase amplitudes $\phi_\mathrm{max}$ listed on the plot. The WG configurations are illustrated on top of the image. (c) Measured average transmission $\bar T$ as a function of the rf voltage $V_\mathrm{rf}$ applied to the IDTs for devices with different number of arms for light with wavelengths $\lambda_{}=910$ ($N_{p}=2$), 890 ($N_{p}=4$), and 900~nm ($N_{p}=6$). The upper horizontal scale displays the corresponding phase modulations. The dashed lines display the calculated transmission according to~\ref{EqAOMID}. Adapted from~\cite{doi:10.1063/1.2768889}.
}
	\label{AOMID}
\end{figure*}

Acousto-optic multiple interference devices (AOMID), first proposed by Beck~\emph{et al.}~\cite{1742-6596-92-1-012006,doi:10.1063/1.2768889,doi:10.1063/1.2821306}, are a class of structures based on the simultaneous modulation of several MZI arms by a single SAW beam. They represent a generalization of the SAW-driven MZI reported in~\cite{doi:10.1063/1.2354411} to implement important acousto-optical functionalities such as ON/OFF switching, harmonic generation, or pulse shapers for the generation of extremely short light pulses. The authors reported two different layouts, namely parallel (P) and series (S) configurations. The parallel configuration is illustrated schematically in Figure~\ref{AOMID}(a). Here, a number $N_{p}\geq2$ of WG arms are connected in parallel and are aligned perpendicularly to the SAW propagation direction. The lateral separation between the WG centers  is chosen so as to yield  SAW-induced phase shifts (as given by~\ref{Optical_phase}) differing by an integer multiple of $2\pi/N_{p}$. In the S configuration, $N_{s}\geq2$ devices in parallel are cascaded and arranged so that the WG arms experience SAW-induced phase shifts differing by an integer multiple of $2\pi/\left(N_{s}N_{p}\right)$. The authors demonstrated that both configurations are feasible to design high-contrast ON/OFF switches. 

We illustrate the operation principle of ON/OFF interference switches using the $N_p=4$ parallel AOMID illustrated in Fig.~\ref{AOMID}(a). The optical transmission intensity $T$ for these structures can be written as a function of the transmission amplitude $t_i$ through each of the WGs according to~\cite{doi:10.1063/1.2821306}:
\begin{equation}
T = \frac{1}{N_p} \left| 	\sum_j^{N_p} {\vec{t}_j(t)}  \right|^2, 
\label{EqAOMID}
\end{equation}  
where $\vec{t}_j\left(t\right)={e^{i\Delta\Phi_j\left(t\right)}}$ and
\begin{equation}
\Delta\Phi_j\left(t\right)=\phi_\mathrm{max}\cos{\left[\omega_\mathrm{SAW}t+i\frac{2\pi}{N_p}\right]}. \nonumber
\end{equation}

Here, $\Delta\Phi_j\left(t\right)$ is the instantaneous optical phase modulation induced by the SAW in the $j^{th}$ WG, which is assumed to have a maximum value $\phi_\mathrm{max}$. The sum in~\ref{EqAOMID} can be schematically visualized by representing the transmission amplitudes $t_j\left(t\right)$ as unity vectors with a phase angle $\Delta\Phi_j\left(t\right)$. In the absence of acoustic excitation, all $\Delta\Phi_j\left(t\right)$ are equal leading to a total intensity transmission $T=1$, as illustrated in Figure~\ref{AOMID}(a):(i). Under a SAW, the vectors ${\vec{t}_j\left(t\right)}$ are no longer collinear: by properly controlling the SAW amplitude it is possible to achieve a vector sum close to zero for all times, as illustrated in Fig.~\ref{AOMID}(a):(ii). This corresponds to the off-state of the switch: the transmitted optical beams add destructively, thus reducing the transmission and increasing the reflection.

%%%%%%%%%%%%%%%%%%%%%%%%%%%%%%%%%%%%%%%%
%%%%%%%%%%%%%%%%%%%%%%%%%%%%%%%%%%%%%%%

In general, for a fixed SAW-induced phase difference $\phi_\mathrm{max}$, the overall transmission of the devices reduces as the number of modulated WG arms increases. Figure~\ref{AOMID}(b) displays calculated time-resolved transmission traces for parallel AOMIDs with different number of arms. The curves show that  optical switches based on  devices with $N_{p}=8$ can achieve OFF/ON extinction ratios $>40$~dB for a SAW-induced phase of $\phi_\mathrm{max}\sim2.4$~rad and $\sim1.6$~rad, respectively~\cite{doi:10.1063/1.2821306}. Similar extinction ratios can be obtained for series devices. In general, the extinction ratio increases much faster with the number of WG arms in the P configuration as compared to the S configuration. P-type AOMIDs require, however, higher SAW-induced phase shifts $\phi_\mathrm{max}$ as compared with S devices. As an example of the switching operation, Fig.~\ref{AOMID}(c) displays the average optical transmission measured for P devices with $N_{p}=2,~4~\mathrm{and}~6$ arms as a function of the SAW-induced phase shift $\phi_\mathrm{max}$ (upper axis) and the voltage $V_{rf}$ of the driving RF signal (lower axis). The measurements were performed for optical wavelengths $\lambda_{}=910$ ($N_{p}=2$), 890 ($N_{p}=4$), and 900~nm ($N_{p}=6$). Here, it can be seen that the average transmission of the devices decreases as $N_{p}$ increases, achieving an extinction ratio of 12~dB for a SAW-induced phase of $\phi_\mathrm{max}\sim2.4$~rad and $N_{p}=6$~\cite{doi:10.1063/1.2768889}. The measured average transmission is normally much higher than the values predicted for perfect devices (i.e., with identical WG arms), which are indicated by the dashed lines in  Fig.~\ref{AOMID}(c). The discrepancy is attributed to fabrication-related small variations in WG length and length. 

As already mentioned, the devices can also be used for the generation of light modulated at higher harmonics of the SAW frequency $f_\mathrm{SAW}$. Particularly, the generation of the sixth harmonic of $f_\mathrm{SAW}$ in P devices with $N_{p}=3$ for $\phi_\mathrm{max}\sim2.4$~rad is also theoretically demonstrated in~\cite{doi:10.1063/1.2821306}. In the same work, the authors also demonstrated theoretically the generation of six sharp light pulses per SAW period by way of S devices with $N_{p}=2$ and $N_{s}=3$ (i.e. three single MZIs connected in series) for $\phi_\mathrm{max}\sim3.6$~rad. In this case, the results also showed that the increase in $\phi_\mathrm{max}$ caused the time narrowing of the generated light pulses. Considering a similar layout to that in~\cite{1742-6596-92-1-012006,doi:10.1063/1.2768889,doi:10.1063/1.2821306}, Barretto and Hvam~\cite{doi:10.1117/12.853801} proposed the introduction of static phase delays in the WG arms in the form of length and thickness differences to make an optical phase shifter. These static phase delays were chosen to ensure destructive interference of the optical fields without SAW excitation. The authors demonstrated that, due to SAW modulation, the frequency of the resulting output optical field was shifted by an amount equal to $f_\mathrm{SAW}$ with respect to the input field, while any other undesired frequency components were dropped due to the static phase delays.

\subsection{Multiple-output interferometers}

%%%%%%%%%%%%%%%%%%%%%%%%%%%%%%%%%%%%%%%%%%%%%%%%%%%%%%%%%%%%%%%%%%%%%%%%%%%%%%%%%%%%%%%%%%%%%%%%%%%%%%%%%%%%%%%%
\begin{figure*}[t!]
	\centering
		\includegraphics[width=0.90\textwidth]{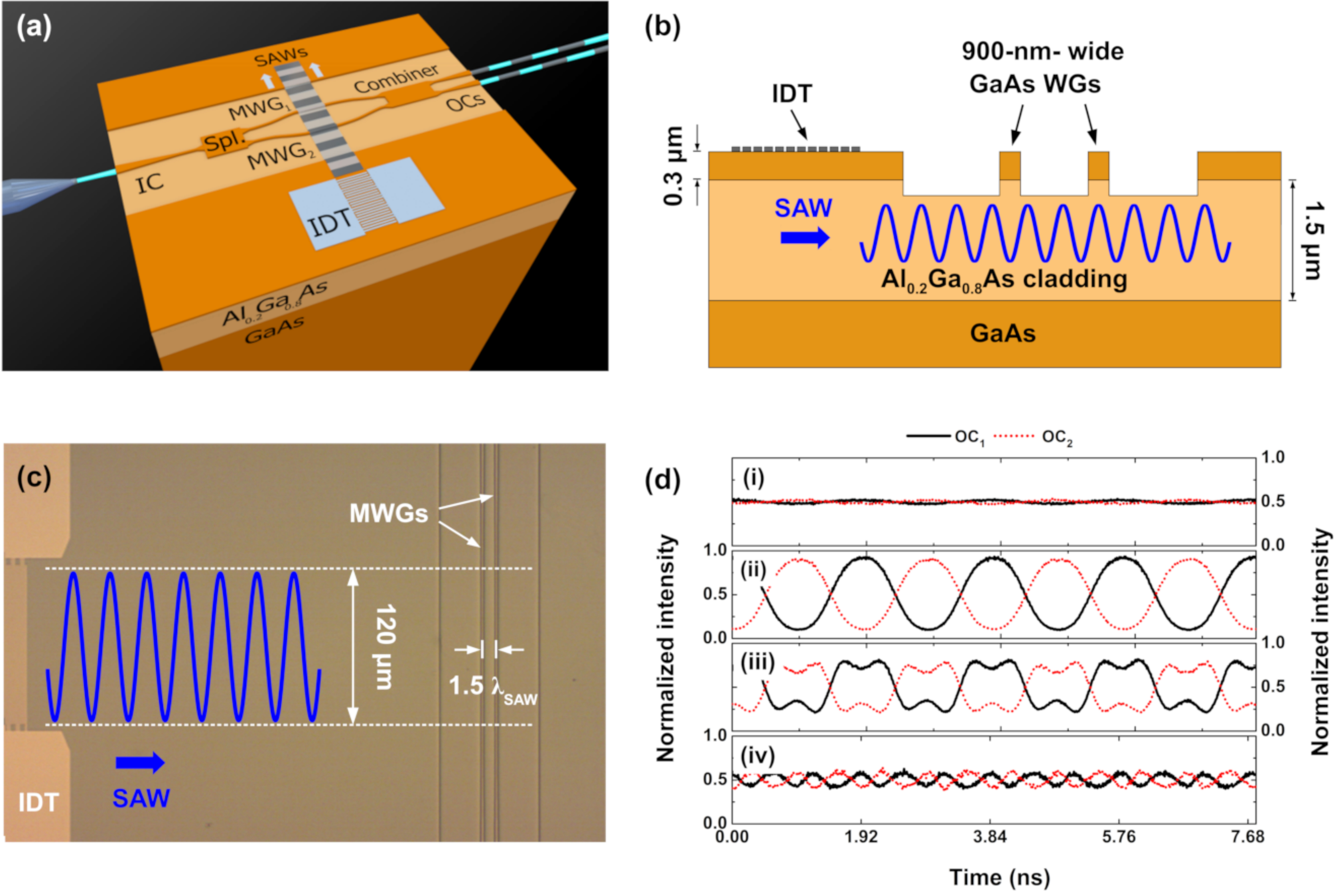}
	\caption{(a) Illustration (not to scale) of the SAW-driven synchronized modulator fabricated on (Al,Ga)As reported in~\cite{Crespo-Poveda:13,doi:10.1117/12.2039538}. By means of a tapered fiber, cw light is coupled into an input channel (IC). The device consists of a splitter and a combiner MMI coupler linked by single mode WGs (MWGs) that are modulated by a SAW generated by an IDT and which propagates perpendicularly to the MGW arms. The light beams leaving the device through both output WGs (OCs) presents a $180^{\circ}$-dephasing synchronization. (b) Corresponding layer cross-section of a device consisting of (Al, Ga)As layers grown by molecular beam epitaxy. (c) Top-view micrograph of one of the fabricated devices with an IDT designed for $\lambda_{\mathrm{SAW}}=5.6~\mu$m with an interaction length $\ell_{int}=120~\mu$m. (d) Time-resolved measurements of the light leaving output WG 1 ($\mathrm{OC_{1}}$, solid line) and output WG 2 ($\mathrm{OC_{2}}$, dotted line) of the same device, recorded for TE polarization and RF powers of (a) $P_{IDT}\approx0$~mW, (b) $P_{IDT}=7$~mW, (c) $P_{IDT}=35$~mW, and (d) $P_{IDT}=49$~mW. The total transmission is normalized to 1. (d) adapted from~\cite{Crespo-Poveda:13}.}
\label{fig:MZI_1X2}
\end{figure*}
%%%%%%%%%%%%%%%%%%%%%%%%%%%%%%%%%%%%%%%%%%%%%%%%%%%%%%%%%%%%%%%%%%%%%%%%%%%%%%%%%%%%%%%%%%%%%%%%%%%%%%%%%%%%%%%%

The acousto-optical devices reported in~\cite{1742-6596-92-1-012006,doi:10.1063/1.2768889,doi:10.1063/1.2821306,doi:10.1117/12.853801} had single input and output WGs. As a result, part of the light transmission was lost due to destructive interference. This issue can be avoided with the addition of extra output WGs to the light combiner so that, at different instants along the SAW period $\left(T_\mathrm{SAW}=2\pi/\omega_\mathrm{SAW}\right)$, the output optical signal switches paths and is selected by a different output WG.

Crespo-Poveda~\emph{et al.}~\cite{Crespo-Poveda:13,doi:10.1117/12.2039538} have reported a MZI modulated by SAWs with a single input WG and two output WGs $\left(1\times2~\mathrm{MZI}\right)$ built upon MMI couplers (MMICs)~\cite{372474} of different width and length [cf. Figure~\ref{fig:MZI_1X2}(a)]. MMICs are optical integrated components comprising a multimode WG with $N$ access singlemode WGs. The number of modes that are excited in the multimode region, each propagating with a different phase constant, depends on the position and the profile of the exciting field. Their interference gives rise to single or multiple images of the input field at different positions perpendicular to the light propagation direction. Therefore, different splitting ratios can be accomplished by an appropriated choice of the input WG position, the input field profile, and the coupling length~\cite{372474}. MMI couplers are extensively used in nowadays integrated photonic applications mainly due to their excellent transmission properties, which enable devices with low losses and good balancing between the output WGs, as well as the possibility of easy design and relaxed fabrication tolerances~\cite{296191}. In the reported devices, the first coupler comprised a balanced $\left(50:50\right)$ splitting ratio 1-to-2-way MMIC that separated the incoming light in two identical beams delivered to two single-mode 900-nm-wide WG arms. The spatial separation of the WG arms was chosen to be $1.5\lambda_\mathrm{SAW}$ (i.e. by an odd multiple -3- of $\lambda_\mathrm{SAW}/2$) to ensure that the arms were modulated with opposite phase by a traveling SAW as in~\cite{doi:10.1063/1.2354411} (cf.~\ref{Neff_change_MZI}). The latter was generated by a regular IDT with a design wavelength of $\lambda_\mathrm{SAW}=5.6~\mu$m (corresponding to $f_\mathrm{SAW}\approx520$~MHz). The modulated WG arms fed a second 2-to-2-way MMIC, which combined the optical fields into the output WGs. The dimensions of the device were optimized for a design light wavelength (i.e., wavelength at which the response of the device is optimal) $\lambda_{0}=900$~nm. The design of the device ensured that when both modulated WG arms deliver light of equal intensity and phase (i.e., without SAW excitation and for several instants along $T_\mathrm{SAW}$ in which the SAW power vanishes), the intensity in both output WGs became also the same. When a SAW-induced phase shift $\left(\Delta\Phi\right)$ was applied between the WG arms, the optical signal could be directed to output WG 1 ($\Delta\Phi=+k\pi$, for $k=1,2,3,...$) or to output WG 2 ($\Delta\Phi=-k\pi$, for $k=1,2,3,...$). 

As in~\cite{doi:10.1063/1.2354411}, the devices were fabricated on an (Al, Ga)As WG sample grown epitaxially on a GaAs $\left(100\right)$ wafer [cf. Figure~\ref{fig:MZI_1X2}(b)]. A top-view micrograph of the modulated region of one of the fabricated devices with an interaction length $\ell_{int}=120~\mu$m is shown in Figure~~\ref{fig:MZI_1X2}(c). The authors characterized the devices for driving powers up to $P_{IDT}=49$~mW, demonstrating strong modulation of the output light beam at the third harmonic of the SAW frequency $f_\mathrm{SAW}$, corresponding to a frequency of $\sim1.5$~GHz. Figure~\ref{fig:MZI_1X2}(d) shows the time-resolved traces obtained for light with TE polarization (similar results were obtained for TM light) for different driving powers $P_{IDT}$. For $P_{IDT}=7$~mW (ii), the modulation reached a peak-to-peak amplitude of approximately 1. After a Fourier analysis of the time-resolved traces, the authors determined the value of the proportionality constant to be $a_{p}=13\times10^{-2}~\mathrm{rad/\sqrt{mW}}$, which is a factor of 2 larger than that obtained in~\cite{doi:10.1063/1.2354411}. A very similar device although with three output WGs has been reported by the same authors in~\cite{crespo2016tunable}. On this occasion, the authors characterized the devices for driving powers of up to $P_{IDT}=108$~mW, for which the optical transmission of the device was entirely dominated by the second harmonic of $f_\mathrm{SAW}$.

\subsection{$N\times N$ modulators driven by surface acoustic waves}\label{N-WGs modulators driven by surface acoustic waves}

\subsubsection{Device layout.}

\begin{figure*}[t!]
	\centering
		\includegraphics[width=0.90\textwidth]{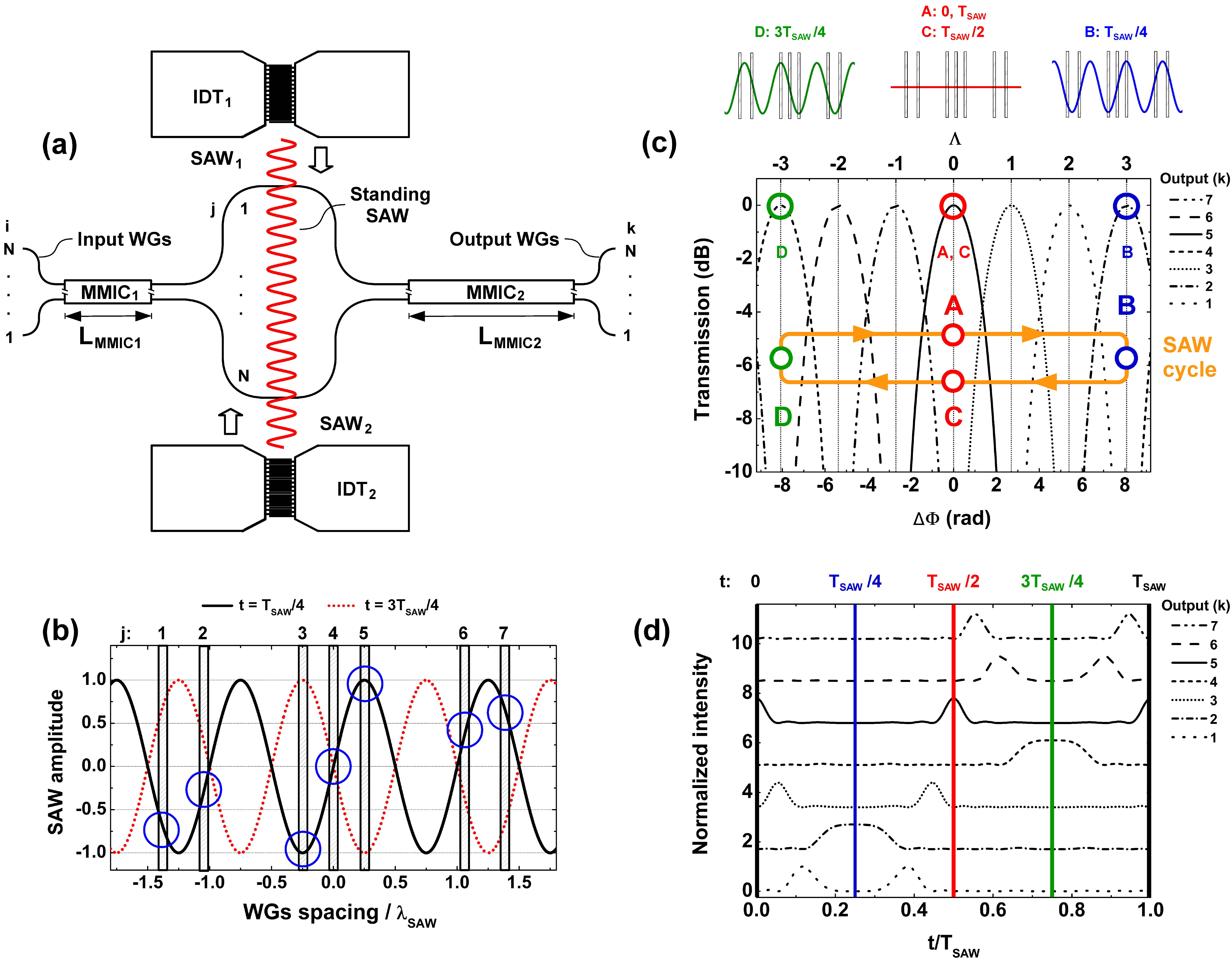}
	\caption{(a) General layout for an acousto-optical modulator with an arbitrary number of access and modulated WGs. For $N\geq3$, the required phase shifts to reconfigure the response of the devices can only be provided by a standing SAW generated by two IDTs. (b) Relative WG spacing (in terms of $\lambda_\mathrm{SAW}$) with respect to the nodes and anti-nodes of the standing SAW, depicted at two different instants along the SAW beating period for a device with seven input and output WGs ($N=7$). The chosen solution corresponds to $\left\langle\kappa_{1},...,\kappa_{7}\right\rangle=\left\langle\mp\sfrac{2}{3},\mp\sfrac{1}{3},\mp1,0,\pm1,\pm\sfrac{1}{3},\pm\sfrac{2}{3}\right\rangle$. The vertical rectangles indicate the WGs positions. (c) Simulated response of the device in (b) as a function of the SAW-induced phase shift $(\Delta\Phi)$ calculated for input WG 3. The closed loop on top represents the effective phase shift $\kappa_{j}\Lambda\delta\Phi^{0}$ (cf.~\ref{Phase_change_arm_j}) on the $j^{th}$ arm at four different instants along the SAW beating period $T_\mathrm{SAW}$. The insets on top display the instantaneous SAW phase with respect to the modulated WGs at different instants of the acoustic period, namely $t=0~(\mathrm{A}),~T_\mathrm{SAW}/4~(\mathrm{B}),~T_\mathrm{SAW}/2~(\mathrm{C}),~\mathrm{and}~3T_\mathrm{SAW}/4~(\mathrm{D})$. (d) Simulated time response of the same device taking $\Lambda=\pm3$ calculated also for input WG 3. The time traces have been vertically displaced for clarity. The vertical lines correspond to the instants labeled as A, B, C, and D in (c).}
\label{fig:MZI_general_layout}
\end{figure*}

The concepts discussed in the previous section can be generalized for optical modulators with an arbitrary number $N$ of input and output WGs~\cite{Crespo-Poveda:16}. The proposed devices comprise two $N$-to-$N$-way MMICs linked by an array of $N$ single-mode WGs. The first MMIC is a balanced splitting ratio coupler of length $L_{MMIC1}=3L_{\pi}/N$, this being the shortest length at which $N$ self-images of the input field are obtained if no restrictions are imposed on the modal excitation. This MMIC divides the incoming optical signal from any of the $N$ input WGs into $N$ identical optical beams. Here, $L_{\pi}$ is the beating length of the MMIC given by~\cite{372474}:  
\begin{equation}
L_{\pi}=\frac{\pi}{\beta_{0}-\beta_{1}}\approx\frac{4n_{r}W_{\mathrm{eff},0}^{2}}{3\lambda_{0}},
\label{MMI_beating_length} 
\end{equation}
where $\beta_{0}$ and $\beta_{0}$ are the fundamental and first mode propagation constants, respectively, $n_{r}$ is the effective refractive index in the MMIC region, $\lambda_{0}$ is the design light wavelength, and $W_{\mathrm{eff},0}$ is the effective width of the fundamental mode of the light propagating in the MMICs, which takes into account the lateral penetration of the optical fields. The $N$ optical beams are coupled into $N$ WG arms, whose refractive index is modulated by one or several SAW beams propagating perpendicularly to them. As will be shown below, for a number of WG arms $N\geq3$, the required phase shifts to modulate the devices can only be accomplished if a standing SAW is generated by way of two IDTs. A second MMIC with length $L_{MMIC2}=\left(3PL_{\pi}\right)/N$, where $P\geq1$, combines the phase-delayed optical fields into the output WGs. Figure~\ref{fig:MZI_general_layout}(a) shows a sketch of the proposed device layout. The shortest devices are obtained by taking $P=1$. In this case, both the splitter and combiner MMICs have the same coupling length. This kind of configuration has been previously used to route the light in  phased-array WDM applications~\cite{Paiam:97} in which by introducing wavelength dispersion in the WG arms, particular wavelengths can be directed to specific output WGs. However, if no wavelength dispersion is introduced (i.e. the static phase difference between the arms in the array is kept equal to an integer multiple of $2\pi$), in the absence of phase modulation the entire device acts as a balanced splitter that divides the optical signal coming from any of the input WGs into $N$ identical optical beams. The latter situation leads to very poor performance of the devices as modulators, increasing the rejection level as $N$ is increased. Considerably better results can be obtained if $P=N-1$. In this configuration, $L_{MMIC1}+L_{MMIC2}=3L_{\pi}$, which is the distance at which a single mirrored image of the input optical field is created inside the MMIC region~\cite{372474}. Therefore, in the preset configuration (i.e., without SAW modulation), the device acts as a cross coupler, so that the light launched at an input port $i$ is delivered to the output port $k_{preset}=N-i+1$. Therefore, it is possible to take advantage of the fact that the light is naturally focused at this output plane of the devices to derive very simple phase relations between the modulated WG arms, i.e., the $\kappa_{j}$ in~\ref{Neff_change_MZI}, that can be used to reconfigure the output channel distribution of the devices.

%%%%%%%%%%%%%%%%%%%%%%%%%%%%%%%%%%%%%%%%%%%%%%%%%%%%%%%%%%%%%%%%%%%%%%%%%%%%%%%%%%%%%%%%%%%%%%%%%%%%%%%%%%%%%%%%
%%%%%%%%%%%%%%%%%%%%%%%%%%%%%%%%%%%%%%%%%%%%%%%%%%%%%%%%%%%%%%%%%%%%%%%%%%%%%%%%%%%%%%%%%%%%%%%%%%%%%%%%%%%%%%%%

\subsubsection{The $\kappa_{j}$ factors.}

The design of acousto-optical light modulators can be greatly simplified if they are designed to operate with very simple phase relations between the modulated WG arms. For a number $N$ of WG arms, these phase relations are provided by a set of $N$ $\kappa_{j}$ factors, as expressed in~\ref{Neff_change_MZI}. As explained before, the latter are numerical factors that indicate the necessary phase relations between the modulated WG arms to switch the optical signal from one output WG to a different one. Here, we will restrict the discussion to a situation where $-1\leq\kappa_{j}\leq1$. With this choice of $\kappa_{j}$, it is possible to take into account situations where the effective refractive index of the modulated WG arm is alternatively increased $\left(0<\kappa_{j}\leq1\right)$ and decreased $\left(-1\leq\kappa_{j}<0\right)$, as naturally occurs in acousto-optical devices due to the SAW oscillation. The procedure to calculate the $\kappa_{j}$ factors can be found in~\cite{Crespo-Poveda:16,doi:10.1117/12.2208868}. For $N=2$, there are two straightforward solutions: $\left\langle \kappa_{1},\kappa_{2}\right\rangle=\left\langle\mp1,\pm1\right\rangle$, which correspond to the MZIs modulated by SAWs reported in~\cite{doi:10.1063/1.2354411,Crespo-Poveda:13,doi:10.1117/12.2039538}. The solutions for up to $N=7$ can be found in~\cite{Crespo-Poveda:16}.

The values of the coefficients $\kappa_{j}$ are determined in practice by the relative spatial separation between the modulated WG arms, since each WG arm will experience an acoustic phase shift and amplitude determined by its position within the SAW modulation profile. In this way, the different sets of $\kappa_{j}$ factors show that  the appropriate positioning of each of the arms in the array within the SAW profile forces an unequal spacing separation of the WGs for $N>3$. In addition, the relative phase difference between the WG arms for $N\geq3$, as determined by the $\kappa_{j}$ factors, can only be sustained in time if a standing SAW is created by use of two IDTs generating two counter-propagating SAW beams. Due to the periodic nature of the SAWs, the WG arms can be placed at distinct periods of the standing SAW, simultaneously meeting the optical and the acoustic constraints. The physical meaning of the $\kappa_{j}$ factors can be better appreciated in Figure~\ref{fig:MZI_general_layout}(b). This picture shows the relative WG spacing with respect to a standing SAW profile (depicted at two different instants of the SAW period -$T_\mathrm{SAW}$-) for a device with $N=7$ input and output WGs. In this case, the set of $\kappa_{j}$ factors corresponds to the solution $\left\langle\kappa_{1},...,\kappa_{7}\right\rangle=\left\langle\mp\sfrac{2}{3},\mp\sfrac{1}{3},\mp1,0,\pm1,\pm\sfrac{1}{3},\pm\sfrac{2}{3}\right\rangle$ (blue circles). This combination of $\kappa_{j}$ factors means that the SAW-induced phase shift must be maximum but of opposite sign at the WG arms 3 and 5. At the WG arms 1 and 7, the SAW-induced phase shift must be also of opposite sign, but with $\sim66\%$ ($\left|\sfrac{2}{3}\right|$) of the amplitude applied to the WG arms 3 and 5. In a similar way, the SAW-induced phase shift at the WG arms 2 and 6 must be of opposite sign but with $\sim33\%$ ($\left|\sfrac{1}{3}\right|$) of the amplitude applied to the WG arms 3 and 5. Finally, no phase shift must be applied to the WG arm 4, where $\kappa_{j}=0$.

In principle, acousto-optical devices with an arbitrary number $N$ of modulated WG arms are possible. However, as the WG arms have a non-zero width, devices with a large number of WG arms can only be designed at the cost of increasing $\lambda_\mathrm{SAW}$ to keep the refractive index modulation approximately constant across each WG arm. This, however, also  results in a slower time response of the modulator.

\subsubsection{Channel assignment.}

The number of possible sets of $\kappa_{j}$ factors (i.e. solutions) depend strongly on whether $N$ is odd or even~\cite{Crespo-Poveda:16,doi:10.1117/12.2208868}. In general, $N\left(N-1\right)$ sets of $\kappa_{j}$ can be found for $N$ odd. These can be arranged in $\left(N-1\right)$ major sets corresponding to solutions that provide the same output WG reconfiguration. For $N$ even, the number of possible solutions reduces to $N$, which can be arranged in two major sets. In order to calculate the output WG reconfiguration corresponding to each set of $\kappa_{j}$, it is useful to rewrite~\ref{Neff_change_MZI} in terms of the SAW-induced phase shift ($\Delta\Phi$) as~\cite{Crespo-Poveda:16,doi:10.1117/12.2208868}: 
\begin{equation}
\Delta\Phi_j\left(t\right)=\Phi^{0}+\kappa_{j}\Lambda\delta\Phi^{0}\cos\left(\omega_\mathrm{SAW}t\right),
\label{Phase_change_arm_j} 
\end{equation}
for $j=1,...,N$, where $\Phi^{0}=\left(2\pi\ell/\lambda\right)n_\mathrm{eff}^{0}$ is the phase shift introduced in the WG arms due to the optical path $\ell$, and $\kappa_{j}\Lambda\delta\Phi^{0}$ is the amplitude of the SAW-induced phase shift in the $j^{th}$ WG arm. Here, $\Lambda$ is an integer and $\delta\Phi^{0}$ is the minimum SAW-induced phase change induced in the WG arms with $\kappa_{j}=\pm1$ needed to reconfigure the output WG distribution. As already mentioned, for colorless devices (i.e. without wavelength dispersion), the length difference between adjacent WG arms must be chosen so that the resulting optical phase shift is zero or an integer multiple of $2\pi$. This is in contrast to WDM devices, which are described below in Section~\ref{Wavelength-division multiplexers modulated by surface acoustic waves}, where the length difference between adjacent WG arms is carefully adjusted to provide wavelength dispersion at the output WGs. 

Figure~\ref{fig:MZI_general_layout}(c) shows the simulated response of the device with $N=7$ input and output WG described in Figure~\ref{fig:MZI_general_layout}(b), calculated as a function of the SAW-induced phase shift $\left(\Delta\Phi\right)$ for input WG 3. The design of the device ensured that the optical signal was selected by output WG 5 without SAW modulation (i.e., identical $\Phi^{0}$ are considered in every WG arm). When the SAW-induced phase shift reached $\Delta\Phi\approx\pm2.69$~rad (corresponding to $\Lambda=\pm1$), the light periodically switched paths among output WGs 3, 5, and 7. By increasing the applied acoustic power in order to achieve $\Delta\Phi\sim5.39$~rad (corresponding to $\Lambda=\pm2$), output WGs 1 and 6 became also available during the SAW oscillation period. For $\Delta\Phi\sim8.09$~rad, all the output WGs become available, with the optical signal being switched from the preset output WG to the rest of output WGs within the SAW period. The closed loop represents the effective phase shift $\kappa_{j}\Lambda\delta\Phi^{0}$ as given by~\ref{Phase_change_arm_j} on the $j^{th}$ arm at four different instants along the SAW beating period $T_\mathrm{SAW}$. In this way, the effective phase shift at the $j^{th}$ arm is $\Delta\Phi_j=\Phi^{0}$ at $t=0~(A)~\mathrm{or}~T_\mathrm{SAW}/2$~(C), whereas it is $\Delta\Phi_j=\Phi^{0}\pm3\kappa_{j}\delta\Phi^{0}$ at $t=T_\mathrm{SAW}/4~(\mathrm{B})~\mathrm{and}~t=3T_\mathrm{SAW}/4$~(D), respectively. This behaviour can be better understood by looking at the insets on top, which display the instantaneous phase of the standing SAW with respect to the modulated WGs at the instants of the standing SAW period labeled as A, B, C, and D. 

Figure~\ref{fig:MZI_general_layout}(d) shows the simulated response of the same device, calculated for $\Lambda=\pm 3$ and input WG 3. An arbitrary $\pi/2$ phase was subtracted to~\ref{Phase_change_arm_j} so that $t=0$ coincided with an instant in which the device behaved as without SAW modulation, and the time traces have been vertically displaced for clarity. For $t=0$, the light is delivered to the preset output WG $5$. During the first quarter of the period, the light goes successively through the output WGs 5, 3, 1, and 2, as $\Delta\Phi$ increases from 0 to the maximum. Then, for $0.25T_\mathrm{SAW}<t<0.5T_\mathrm{SAW}$ the assignment reverses, going through the output WGs 2, 1, 3, and 5, as $\Delta\Phi$ approaches zero. For the second half of the beating cycle, when $\Delta\Phi$ is negative, the dynamic allocation of the channels obey the following sequence: 5, 7, 6, 4, 6, 7, 5. The vertical lines correspond to the instants $t=0, T_\mathrm{SAW}/4, T_\mathrm{SAW}/2,~\mathrm{and}~3T_\mathrm{SAW}/4$ depicted in Figure~\ref{fig:MZI_general_layout}(c).

\subsection{Wavelength-division multiplexers modulated by surface acoustic waves}\label{Wavelength-division multiplexers modulated by surface acoustic waves}

	%%%%%%%%%%%%%%%%%%%%%%%%%%%%%%%%%%%%%%%%%%%%%%%%%%%%%%%%%%%%%%%%%%%%%%%%%%%%%%%%%%%%%%%%%%%%%%%%%%%%%%%%%%%%%%%%
\begin{figure*}[t!]
\centering
\includegraphics[width=0.90\textwidth]{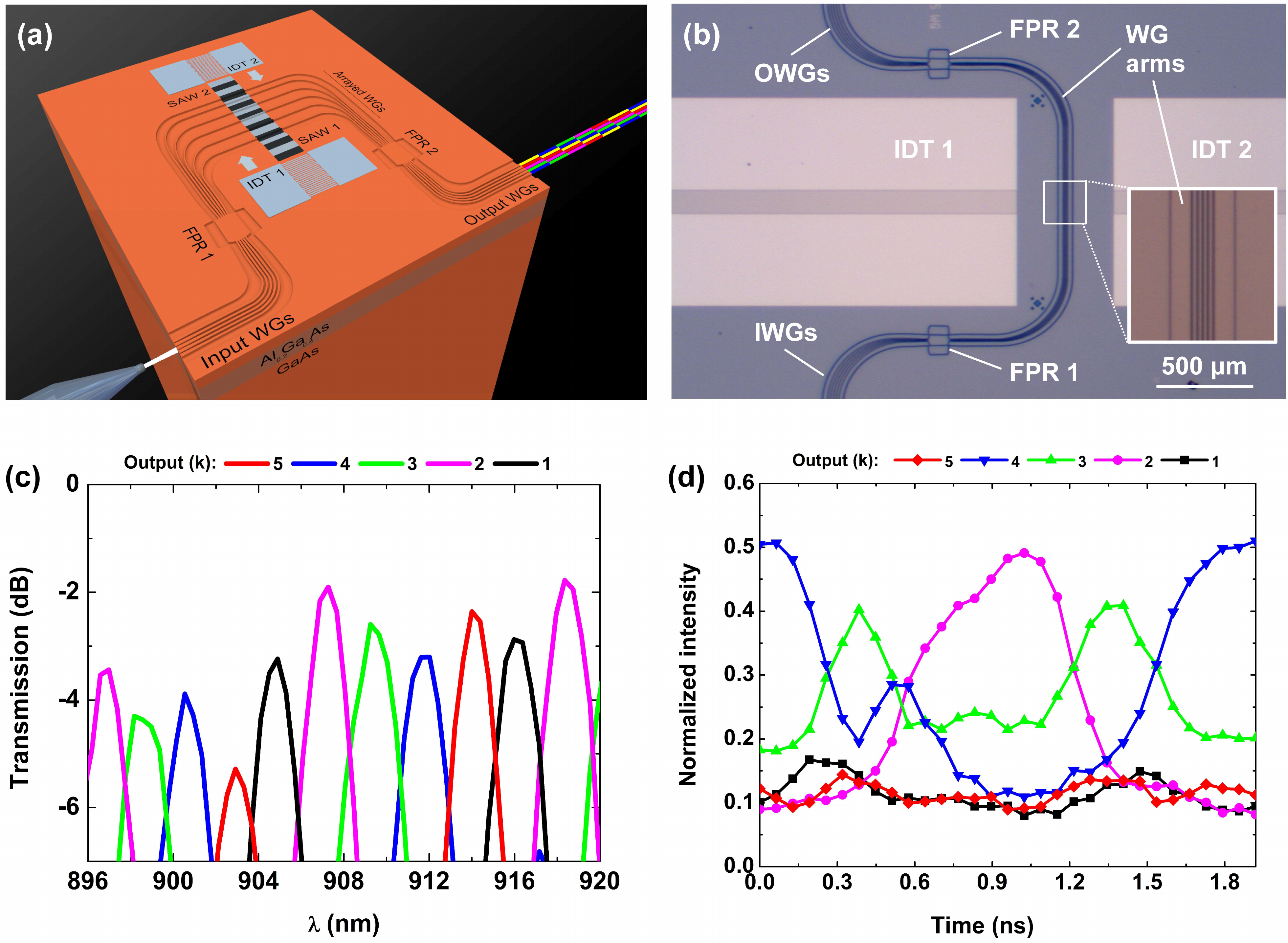}
\caption{(a) Illustration (not to scale) of the FPRs-based WDM devices demonstrated in~\cite{Crespo-Poveda:15}. The light is coupled into an input channel (Input WGs) by means of a optical fibre probe. Next, it is consecutively split and combined by the splitter and combiner couplers (Coupler 1 and Coupler 2, respectively), which are linked by an array of single-mode WGs (Arrayed WGs). These are modulated by a standing SAW generated by two IDTs that dynamically modifies the output channel corresponding to a given optical wavelength. Reproduced from~\cite{Crespo-Poveda:15}. (b) Top view micrograph of one the devices. An enlarged view of the SAW-light interaction region, denoted by the white square, is shown in the inset. (c) Measured spectral response of the same WDM device corresponding to input WG 3 for light with TE polarization. (d) Time-resolved measurements recorded for the light leaving the output WGs of the device, measured for the input WG 3, light wavelength $\lambda=899$~nm, TE polarization, and a driving power of $P_{IDT}=80$~mW on each IDT. The total transmission is normalized to 1. (b), (c), and (d) adapted from~\cite{Crespo-Poveda:15}.}
\label{fig:AWG_results}
\end{figure*}
%%%%%%%%%%%%%%%%%%%%%%%%%%%%%%%%%%%%%%%%%%%%%%%%%%%%%%%%%%%%%%%%%%%%%%%%%%%%%%%%%%%%%%%%%%%%%%%%%%%%%%%%%%%%%%%%

Wavelength multiplexers and demultiplexers are key components in WDM technology. The latter allows for using all the frequency channels within the bandwidth of large-capacity optical fibers by combining, before transmission, different wavelength channels into a single WDM signal. The received WDM signal can be separated using an inverse device (demultiplexer) which splits the incoming signal into its constituting wavelength channels. In these devices, wavelength dispersion is provided by an array of WG arms, with length calculated so that constructive interference is granted at each output WG for a different wavelength. This technology plays a pivotal role in nowadays communications, as it enables very efficient long-distance data transmission. 

The wavelength routing capabilities of WDMs are predetermined during the design stage and cannot be modified during operation. Greater wavelength routing flexibility can thus be achieved if tuning techniques are applied to these devices. In this way, tuning of WDM devices by use of the thermo-optic~\cite{661019,681462,Xiao:04} or electro-optic~\cite{5588617,6260063} effects has already been demonstrated. In this section, we review the most important results corresponding to the modulation of WDM devices by use of piezoelectrically-generated SAWs reported in~\cite{doi:10.1117/12.2208868,Crespo-Poveda:15}. The authors demonstrated two different architectures for tunable WDM devices designed to operate with five equally-spaced wavelength channels. One layout used MMICs as power splitters and combiners, whereas free propagation regions~\cite{84502,372441} (FPRs) were employed in the other layout. The spectral properties of the resulting devices are strongly influenced by the choice of couplers. In both cases, the optical properties of the WG arms were modulated by a standing SAW generated by two IDTs, and the dynamic allocation of the wavelength channels with modulation rates in the low GHz frequency range was demonstrated.

Wavelength multiplexers using N-to-N-way MMICs as power splitters/combiners are possible by following a similar configuration to that described in Section~\ref{N-WGs modulators driven by surface acoustic waves} and by choosing the length of the $j^{th}$ WG arm $\left(\ell_{j}\right)$ to be~\cite{doi:10.1117/12.2208868}: 
\begin{equation}
\ell_{j}=\ell_{0}+\alpha_{j}\Delta\ell
\label{Length_difference_arm_j_AWG_general} 
\end{equation}
for $j=1,...,N$, where $\ell_{0}$ is the length of the shortest reference WG arm, $\alpha_{j}$ are integer factors, and $\Delta\ell$ is the length difference that introduces a phase shift of $2\pi/N$ between adjacent wavelength channels. The latter can be explicitly written as~\cite{doi:10.1117/12.2208868,Crespo-Poveda:15,Paiam:97}: 
\begin{equation}
\Delta\ell=\frac{\lambda_{0}\left(\lambda_{0}+\Delta\lambda_{0}\right)}{N\Delta\lambda\left[n_\mathrm{eff}^{0}-\lambda_{0}\left(dn_\mathrm{eff}/d\lambda\right)\right]}
\label{Length_difference_expression} 
\end{equation}
where $\lambda_{0}$ is the design light wavelength, $\Delta\lambda$ is the spacing between adjacent wavelength channels, and $dn_\mathrm{eff}/d\lambda$ is the derivative of the effective refractive index with respect to the optical wavelength $\lambda$. In general, because of the transfer phases between the input and output WGs in MMICs~\cite{372474}, the optical length of each WG arm does not increase linearly with respect to the preceding WG arm. This has important consequences for the channel assignment, since  adjacent wavelength channels are not necessarily routed towards adjacent output WGs.

The WDM devices modulated by SAWs demonstrated in~\cite{doi:10.1117/12.2208868,Crespo-Poveda:15} comprised two 5-to-5-way MMICs of length $L_{c}=3L_{\pi}/5$ linked by an array of $N=5$ single-mode WG arms, following the configuration demonstrated by Paiam and MacDonald~\cite{Paiam:97}. The latter, which leads to considerably shorter WDM devices, is possible if~\ref{Length_difference_arm_j_AWG_general} is modified according to: 
\begin{equation}
\ell_{j}=\ell_{0}+\left(\alpha_{j}+\delta\alpha_{j}\right)\Delta\ell,
\label{Length_difference_arm_j_AWG_general_mod} 
\end{equation}
where $\delta\alpha_{j}\in\boldsymbol{R}$ are corrections introduced to the $\alpha_{j}$ factors, which are therefore no longer integers. The authors calculated $\Delta\ell=19.26~\mu$m for a design optical wavelength $\lambda_{0}=910$~nm, TE polarization, and a spacing between the different wavelength channels $\Delta\lambda=\pm2$~nm. The calculated $\alpha_{j}$ factors as expressed in~\ref{Length_difference_arm_j_AWG_general_mod} were $\left\langle\alpha_{1}+\delta\alpha_{1},...,\alpha_{5}+\delta\alpha_{5}\right\rangle\approx\left\langle0,1.99,2.98,4.04,5.98\right\rangle$. Two different solutions of $\kappa_{j}$ factors were essayed, in particular $\left\langle\kappa_{1},...,\kappa_{5}\right\rangle=\left\langle\mp\sfrac{1}{2},\mp1,0,\pm1,\pm\sfrac{1}{2}\right\rangle$ and $\left\langle\mp1,\pm\sfrac{1}{2},0,\mp\sfrac{1}{2},\pm1\right\rangle$. In order to satisfy the optical constraints (given by~\ref{Length_difference_arm_j_AWG_general_mod}) and the acoustic constraints [cf. Figure~\ref{fig:MZI_general_layout}(b)], the WG arms were placed at the appropriated positions by means of large-radius S-bends. Two IDTs were employed to modulate the WG arms. In this configuration, two counter-propagating traveling SAWs interfered to create a standing SAW. %The IDTs, with an aperture of $120~\mu$m, were designed for an operation wavelength of $\lambda_\mathrm{SAW}=5.6~\mu$m (corresponding to $f_\mathrm{SAW}=520$~MHz). 
The WDM devices were fabricated in a WG sample grown on $\left(100\right)$ GaAs wafer similar to that employed in~\cite{doi:10.1063/1.2354411,Crespo-Poveda:13}. With no RF power applied to the IDTs, each wavelength signal mainly coupled to the output WG previously determined by the design of the multiplexer. By increasing $P_{IDT}$, the different wavelength channels could be sequentially assigned to each output WG of the device with rate defined by $f_\mathrm{SAW}$.
  
The operation principle is similar in the second layout built upon FPRs [cf.~Figure~\ref{fig:AWG_results}(a)]. The devices comprised two FPRs with the same focal length where light diffracts, linked by an array of single-mode WG arms. In this configuration, the optical length of the $j^{th}$ arrayed WG arm $\left(\ell_{j}\right)$ linearly increases by a fixed amount with respect to the preceding WG arm according to the following expression: 
\begin{equation}
\ell_{j}=\ell_{1}+\left(j-1\right)\Delta\ell, 
\label{Length_difference_arm_j_AWG_FPR} 
\end{equation}
for $j=1,...,N$, where $\ell_{1}$ is the length of the shortest reference WG arm (which corresponds to the WG arm 1), and $\Delta\ell=(m\lambda_{0}/n_\mathrm{eff}^{0})$ is the introduced length difference, with $m$ an integer. This length difference causes a wavelength-dependent phase change which varies linearly along the output plane of the arrayed WG arms. As a result, constructive interference will occur for different wavelengths at different points along the focal plane of the second FPR, where the output WGs are placed. As the phase change varies linearly, neighboring wavelength channels will therefore be selected by adjacent output WGs. This property also implies that the devices can only be properly tuned if the SAW-induced phase difference increases linearly between adjacent WG arms. 

The devices demonstrated by the authors were optimized for a design wavelength $\lambda_{0}=900$~nm and contained an array of $N=5$ WG arms, requiring a set of $\kappa_{j}$ factors given by $\left\langle\kappa_{1},...,\kappa_{5}\right\rangle=\left\langle\mp1,\mp\sfrac{1}{2},0,\pm\sfrac{1}{2},\pm1\right\rangle$. The devices were processed on the same wafer as the WDM devices built upon MMICs. The required phase shifts to modulate the 900-nm-wide WG arms were also provided by a standing SAW generated by two IDTs, identical to those described above. These are clearly visible in the top view micrograph of Figure~\ref{fig:AWG_results}(b). Figure~\ref{fig:AWG_results}(c) shows the measured spectral response of the same device, corresponding to input WG 3 and TE polarization. Here, it can be clearly seen that different wavelength channels interfere constructively at different output WGs. The time response of the device can be seen in Figure~\ref{fig:AWG_results}(d), which shows the time-resolved traces recorded for input WG 3 along one acoustic period, for light with TE polarization and $\lambda_{}=899$~nm. A total driving power of $P_{IDT}=80$~mW was applied on each IDT. The total transmission is normalized to 1. Here, the light oscillated between output WG 3 and adjacent output WGs 2 and 4 during a SAW cycle. The traces corresponding to the two latter output WGs were dephased by $180^{\circ}$ with respect to each other. In contrast, two maxima in the transmission were visible at output WG 3 corresponding to the times in which the standing SAW vanishes. A negligible modulation was also observed at the outer output WGs 1 and 5.

\section{Conclusions and perspectives}\label{Conclusions and perspectives}

Throughout this paper, we have classified and reviewed the different developments reported during the past years related to the modulation of optical WG devices by SAWs. First, the fundamentals of optical waveguiding in 3D structures was given in Section~\ref{Overview of optical waveguiding}. Next, the spatial distribution of SAW fields in semiconductors as well as the different methods to excite them by use of monolithic structures have been reviewed in Section~\ref{Surface acoustic waves in piezoelectric semiconductors}. In Section~\ref{Acousto-optical interaction in semiconductor structures}, we have discussed the effects of the dynamic SAW strain on the optical properties of semiconductors and described how the interaction between SAWs and optical fields in 3D optical WGs can be optimized to design efficient acousto-optical devices. Next, we have gathered and classified some of the most important developments regarding the modulation of different types of optical WG devices by means of SAWs. In particular, we have reviewed tunable ring modulators and the interesting concept of sub-optical wavelength modulation (Section~\ref{Ring resonators modulated by surface acoustic waves}), as well as Mach-Zehnder interferometers and multiplexers built upon ridge or rib WGs (Section~\ref{Interferometric acousto-optic devices}). All these devices offer an excellent platform for on-chip modulation of optical signals. 

Photonic crystals are an interesting alternative to 3D optical waveguides that can also be combined with them in monolithic optical circuits. In these structures, periodicity gives rise to a photonic band gap, which prevents light propagation within a certain frequency range. By breaking periodicity through the introduction of a line defect, an optical WG can be created where photons of a given frequency can be guided. If a cavity layer is placed between two distributed Bragg reflectors or a point defect is introduced into a 2D photonic crystal, an optical cavity is created where photons of a given frequency can be confined. This extra confinement gives rise to an enhanced interaction between the optical and the acoustic fields when the photonic crystal is tuned by use of piezoelectrically generated SAWs. A thorough study of SAW tuning of optical cavities in 2D photonic crystals has been done by Fuhrmann~\emph{et al.}~\cite{fuhrmann2011dynamic,fuhrmann2014high}. Moreover, SAW tuning provides an efficient way for fine control of light-matter coupling when low-dimensional excitonic structures, such as quantum dots or quantum wells are embedded inside the optical cavity, paving the way towards important solid state functionalities. As an example, the acoustic manipulation using SAWs of polaritons and its quantum condensates in optical microcavities with embedded quantum wells, which merge when strong coupling conditions are fulfilled, has the potential to yield important breakthroughs in the field of quantum simulation. A very recent review has been done by Cerda-M\'{e}ndez~\emph{et al.} in~\cite{0022-3727-51-3-033001}.

Currently, there is an ongoing research effort from research groups in Europe, Asia, and North America as part of the SAWtrain consortium~\cite{DBMarieCuriewebsite}, which covers a great variety of topics from advanced acousto-optic waveguide modulators operating at telecommunication wavelengths to acoustic transport in graphene or SAWs for quantum control of device structures. For all these reasons, everything suggests that the modulation of photonic structures by SAWs is going to remain the objective of intense research. In this way, integrated devices of increasing complexity simultaneously combining different types of optical waveguiding and confinement as well as different device functionalities should be expected in the coming years.

\section*{Acknowledgments}

The authors acknowledge Alberto Hern\'{a}ndez-M\'{i}nguez for helpful discussions. This work has received funding from the European Union’s Horizon 2020 research and innovation programme under the Marie Sk\l{}odowska-Curie grant agreement No. 642688 (SAWtrain).

\bibliographystyle{unsrt}
\bibliography{literature}

\begin{thebibliography}{100}

\bibitem{Pruessner:07}
M.~W. Pruessner, T.~H. Stievater, M.~S. Ferraro, and W.~S. Rabinovich.
\newblock Thermo-optic tuning and switching in $\mathrm{SOI}$ waveguide
  fabry-perot microcavities.
\newblock {\em Opt. Express}, 15(12):7557--7563, 2007.

\bibitem{Li:14}
X.~Li, H.~Xu, X.~Xiao, Z.~Li, Y.~Yu, and J.~Yu.
\newblock Fast and efficient silicon thermo-optic switching based on reverse
  breakdown of pn junction.
\newblock {\em Opt. Lett.}, 39(4):751--753, 2014.

\bibitem{Yang:11}
M.~Yang, W.~M.~J. Green, S.~Assefa, J.~Van Campenhout, B.~G. Lee, C.~V. Jahnes,
  F.~E. Doany, C.~L. Schow, J.~A. Kash, and Y.~A. Vlasov.
\newblock Non-blocking $\mathrm{4\times 4}$ electro-optic silicon switch for
  on-chip photonic networks.
\newblock {\em Opt. Express}, 19(1):47--54, 2011.

\bibitem{Xing:13}
J.~Xing, Z.~Li, Y.~Yu, and J.~Yu.
\newblock Low cross-talk $\mathrm{2\times 2}$ silicon electro-optic switch
  matrix with a double-gate configuration.
\newblock {\em Opt. Lett.}, 38(22):4774--4776, 2013.

\bibitem{Lu:11}
H.~Lu, X.~Liu, L.~Wang, Y.~Gong, and D.~Mao.
\newblock Ultrafast all-optical switching in nanoplasmonic waveguide with kerr
  nonlinear resonator.
\newblock {\em Opt. Express}, 19(4):2910--2915, 2011.

\bibitem{doi:10.1021/nl404356t}
W.~Li, B.~Chen, C.~Meng, W.~Fang, Y.~Xiao, X.~Li, Z.~Hu, Y.~Xu, L.~Tong,
  H.~Wang, W.~Liu, J.~Bao, and Y.~R. Shen.
\newblock Ultrafast all-optical graphene modulator.
\newblock {\em Nano Lett.}, 14(2):955--959, 2014.

\bibitem{almeida2004all}
V.~R. Almeida, C.~A. Barrios, R.~R. Panepucci, and M.~Lipson.
\newblock All-optical control of light on a silicon chip.
\newblock {\em Nature}, 431(7012):1081--1084, 2004.

\bibitem{1159054}
T.~A. Ibrahim, W.~Cao, Y.~Kim, J.~Li, J.~Goldhar, P.~. Ho, and C.~H. Lee.
\newblock All-optical switching in a laterally coupled microring resonator by
  carrier injection.
\newblock {\em IEEE Photon. Technol. Lett.}, 15(1):36--38, 2003.

\bibitem{1386333}
T.~Sadagopan, S.~J. Choi, Sang~Jun Choi, K.~Djordjev, and P.~D. Dapkus.
\newblock Carrier-induced refractive index changes in $\mathrm{InP}$-based
  circular microresonators for low-voltage high-speed modulation.
\newblock {\em IEEE Photon. Technol. Lett.}, 17(2):414--416, 2005.

\bibitem{1073206}
R.~Soref and B.~Bennett.
\newblock Electrooptical effects in silicon.
\newblock {\em IEEE J. Quantum Electron.}, 23(1):123--129, 1987.

\bibitem{doi:10.1063/1.1371786}
S.~Kim and V.~Gopalan.
\newblock Strain-tunable photonic band gap crystals.
\newblock {\em Appl. Phys. Lett.}, 78(20):3015--3017, 2001.

\bibitem{doi:10.1063/1.1649803}
C.~W. Wong, P.~T. Rakich, S.~G. Johnson, M.~Qi, H.~I. Smith, E.~P. Ippen, L.~C.
  Kimerling, Y.~Jeon, G.~Barbastathis, and S.~G. Kim.
\newblock Strain-tunable silicon photonic band gap microcavities in optical
  waveguides.
\newblock {\em Appl. Phys. Lett.}, 84(8):1242--1244, 2004.

\bibitem{doi:10.1063/1.1754276}
R.~M. White and F.~W. Voltmer.
\newblock Direct piezoelectric coupling to surface acoustic waves.
\newblock {\em Appl. Phys. Lett.}, 7(12):314--316, 1965.

\bibitem{1126862}
R.~Shubert and J.~H. Harris.
\newblock Optical surface waves on thin films and their application to
  integrated data processors.
\newblock {\em IEEE Trans. Microw. Theory Techn.}, 16(12):1048--1054, 1968.

\bibitem{1076309}
D.~Maydan.
\newblock Acousto-optical pulse modulators.
\newblock {\em IEEE J. Quantum Electron.}, 6(1):15--24, 1970.

\bibitem{doi:10.1063/1.1654747}
M.~L. Shah.
\newblock Fast acoustic diffraction-type optical waveguide modulator.
\newblock {\em Appl. Phys. Lett.}, 23(10):556--558, 1973.

\bibitem{doi:10.1063/1.106625}
C.~S. Tsai and P.~Le.
\newblock $4\times 4$ nonblocking integrated acousto-optic space switch.
\newblock {\em Appl. Phys. Lett.}, 60(4):431--433, 1992.

\bibitem{145254}
A.~Kar-Roy and C.~S. Tsai.
\newblock $\mathrm{8\times 8}$ symmetric nonblocking integrated acoustooptic
  space switch module on $\mathrm{LiNbO_{3}}$.
\newblock {\em IEEE Photonics Technology Letters}, 4(7):731--734, 1992.

\bibitem{tsai2013guided}
C.~S. Tsai~(Ed.).
\newblock {\em Guided-wave acousto-optics: interactions, devices, and
  applications}.
\newblock Berlin: Springer, 1990.

\bibitem{Goell:73}
J.~E. Goell.
\newblock Rib waveguide for integrated optical circuits.
\newblock {\em Appl. Opt.}, 12(12):2797--2798, 1973.

\bibitem{doi:10.1063/1.88279}
Y.~Ohmachi and J.~Noda.
\newblock Electro-optic light modulator with branched ridge waveguide.
\newblock {\em Appl. Phys. Lett.}, 27(10):544--546, 1975.

\bibitem{doi:10.1063/1.87992}
W.~E. Martin.
\newblock A new waveguide switch/modulator for integrated optics.
\newblock {\em Appl. Phys. Lett.}, 26(10):562--564, 1975.

\bibitem{390234}
K.~Noguchi, O.~Mitomi, H.~Miyazawa, and S.~Seki.
\newblock A broadband $\mathrm{Ti:LiNbO_{3}}$ optical modulator with a ridge
  structure.
\newblock {\em J. Lightwave Technol.}, 13(6):1164--1168, 1995.

\bibitem{250348}
A.~Salokatve, H.~Jeon, J.~Ding, M.~Hovinen, A.~V. Nurmikko, D.~C. Grillo,
  Li~He, J.~Han, Y.~Fan, M.~Ringle, R.~L. Gunshor, G.~C. Hua, and N.~Otsuka.
\newblock Continuous-wave, room temperature, ridge waveguide green-blue diode
  laser.
\newblock {\em Electron. Lett.}, 29(25):2192--2194, 1993.

\bibitem{849055}
B.~Borchert, A.~Y. Egorov, S.~Illek, and H.~Riechert.
\newblock Static and dynamic characteristics of 1.29-$\mu$m gainnas
  ridge-waveguide laser diodes.
\newblock {\em IEEE Photon. Technol. Lett.}, 12(6):597--599, 2000.

\bibitem{ebeling1993integrated}
K.~J. Ebeling.
\newblock {\em Integrated optoelectronics: waveguide optics, photonics,
  semiconductors}.
\newblock Berlin: Springer, 1 edition, 1993.

\bibitem{book:166107}
R.~G. Hunsperger.
\newblock {\em Integrated optics: theory and technology}.
\newblock New York: Springer, 6 edition, 2009.

\bibitem{1017609}
K.~Saitoh and M.~Koshiba.
\newblock Full-vectorial imaginary-distance beam propagation method based on a
  finite element scheme: application to photonic crystal fibers.
\newblock {\em IEEE J. Quantum Electron.}, 38(7):927--933, 2002.

\bibitem{oliner1994acoustic}
A.~A. Oliner~(Ed.).
\newblock {\em Acoustic surface waves}.
\newblock Berlin: Springer, 1994.

\bibitem{Ash1985}
E.~A. Ash and E.~G.~S. Paige~(Eds.).
\newblock {\em Rayleigh-Wave Theory and Application}.
\newblock Berlin: Springer, 1985.

\bibitem{1449830}
R.~M. White.
\newblock Surface elastic waves.
\newblock {\em Proc. IEEE}, 58(8):1238--1276, 1970.

\bibitem{royer2000elastic}
D.~Royer and E.~Dieulesaint.
\newblock {\em Elastic waves in solids I: free and guided propagation}.
\newblock Berlin: Springer, 2000.

\bibitem{PhysRevB.40.7874}
A.~Wixforth, J.~Scriba, M.~Wassermeier, J.~P. Kotthaus, G.~Weimann, and
  W.~Schlapp.
\newblock Surface acoustic waves on $\mathrm{GaAs/Al_{x}Ga_{1-x}As}$
  heterostructures.
\newblock {\em Phys. Rev. B}, 40:7874--7887, 1989.

\bibitem{doi:10.1063/1.350747}
A.~Neubrand and P.~Hess.
\newblock Laser generation and detection of surface acoustic waves: Elastic
  properties of surface layers.
\newblock {\em J. Appl. Phys.}, 71(1):227--238, 1992.

\bibitem{campbell1989surface}
C.~Campbell.
\newblock {\em Surface acoustic wave devices and their signal processing
  applications}.
\newblock London: Academic Press, 1989.

\bibitem{royer2000elasticII}
D.~Royer and E.~Dieulesaint.
\newblock {\em Elastic waves in solids II: generation, acousto-optic
  interaction, applications}.
\newblock Berlin: Springer, 2000.

\bibitem{hashimoto2000surface}
K.~Hashimoto.
\newblock {\em Surface acoustic wave devices in telecommunications: modelling
  and simulation}.
\newblock Berlin: Springer, 2000.

\bibitem{doi:10.1063/1.1625419}
M.~M. de~Lima~Jr., F.~Alsina, W.~Seidel, and P.~V. Santos.
\newblock Focusing of surface-acoustic-wave fields on (100) $\mathrm{GaAs}$
  surfaces.
\newblock {\em J. Appl. Phys.}, 94(12):7848--7855, 2003.

\bibitem{4249186}
K.~Yamanouchi and H.~Furuyashiki.
\newblock New low-loss $\mathrm{SAW}$ filter using internal floating electrode
  reflection types of single-phase unidirectional transducer.
\newblock {\em Electron. Lett.}, 20(24):989--990, 1984.

\bibitem{1539925}
C.~K. Campbell and C.~B. Saw.
\newblock Analysis and design of low-loss saw filters using single-phase
  unidirectional transducers.
\newblock {\em IEEE Trans. Ultrason., Ferroelect., Freq. Control},
  34(3):357--367, 1987.

\bibitem{171324}
T.~Thorvaldsson and B.~P. Abbott.
\newblock Low loss $\mathrm{SAW}$ filters utilizing the natural single phase
  unidirectional transducer ($\mathrm{NSPUDT}$).
\newblock In {\em IEEE Symposium on Ultrasonics}, pages 43--48 vol.1, 1990.

\bibitem{0964-1726-15-6-003}
W.~Wang, S.~He, S.~Li, and Y.~Pan.
\newblock High frequency stability oscillator for surface acoustic wave-based
  gas sensor.
\newblock {\em Smart Mater. Struct.}, 15(6):1525, 2006.

\bibitem{1347-4065-47-5S-4065}
J.~Kondoh, Y.~Okiyama, S.~Mikuni, Y.~Matsui, M.~Nara, T.~Mori, and H.~Yatsuda.
\newblock Development of a shear horizontal surface acoustic wave sensor system
  for liquids with a floating electrode unidirectional transducer.
\newblock {\em Jpn. J. Appl. Phys.}, 47(5S):4065, 2008.

\bibitem{doi:10.1063/1.4975803}
Maria~K. Ekstr\"{o}m, T.~Aref, J.~Runeson, J.~Bj\"{o}rck, I.~Bostr\"{o}m, and
  P.~Delsing.
\newblock Surface acoustic wave unidirectional transducers for quantum
  applications.
\newblock {\em Appl. Phys. Lett.}, 110(7):073105, 2017.

\bibitem{doi:10.1063/1.2976135}
J.~R. Gell, M.~B. Ward, R.~J. Young, R.~M. Stevenson, P.~Atkinson, D.~Anderson,
  G.~A.~C. Jones, D.~A. Ritchie, and A.~J. Shields.
\newblock Modulation of single quantum dot energy levels by a
  surface-acoustic-wave.
\newblock {\em Appl. Phys. Lett.}, 93(8):081115, 2008.

\bibitem{doi:10.1021/nl203461m}
A.~Hern\'{a}ndez-M\'{i}nguez, M.~M\"{o}ller, S.~Breuer, C.~Pf\"{u}ller,
  C.~Somaschini, S.~Lazi\'{c}, O.~Brandt, A.~Garc\'{i}a-Crist\'{o}bal, M.~M.
  de~Lima, A.~Cantarero, L.~Geelhaar, H.~Riechert, and P.~V. Santos.
\newblock Acoustically driven photon antibunching in nanowires.
\newblock {\em Nano Lett.}, 12(1):252--258, 2012.
\newblock PMID: 22142481.

\bibitem{doi:10.1021/nl1042775}
J.~B. Kinzel, D.~Rudolph, M.~Bichler, G.~Abstreiter, J.~J. Finley,
  G.~Koblm\"{u}ller, A.~Wixforth, and H.~J. Krenner.
\newblock Directional and dynamic modulation of the optical emission of an
  individual gaas nanowire using surface acoustic waves.
\newblock {\em Nano Lett.}, 11(4):1512--1517, 2011.
\newblock PMID: 21355606.

\bibitem{doi:10.1063/1.336130}
J.~Z. Wilcox and R.~E. Brooks.
\newblock Time-fourier transform by a focusing array of phased acoustic wave
  transducers.
\newblock {\em J. Appl. Phys.}, 58(3):1148--1159, 1985.

\bibitem{doi:10.1063/1.336131}
J.~Z. Wilcox and R.~E. Brooks.
\newblock Frequency-dependent beam steering by a focusing array of surface
  acoustic wave transducers: Experiment.
\newblock {\em J. Appl. Phys.}, 58(3):1160--1168, 1985.

\bibitem{1509798}
T.-T. Wu, H.-T. Tang, Y.-Y. Chen, and P.-L. Liu.
\newblock Analysis and design of focused interdigital transducers.
\newblock {\em IEEE Trans. Ultrason., Ferroelect., Freq. Control},
  52(8):1384--1392, 2005.

\bibitem{doi:10.1063/1.3190518}
D.~Yudistira, S.~Benchabane, D.~Janner, and V.~Pruneri.
\newblock Surface acoustic wave generation in $\mathrm{ZX}$-cut
  $\mathrm{LiNbO_{3}}$ superlattices using coplanar electrodes.
\newblock {\em Appl. Phys. Lett.}, 95(5):052901, 2009.

\bibitem{doi:10.1063/1.3599569}
D.~Yudistira, S.~Benchabane, D.~Janner, and V.~Pruneri.
\newblock Diffraction less and strongly confined surface acoustic waves in
  domain inverted $\mathrm{LiNbO_{3}}$ superlattices.
\newblock {\em Appl. Phys. Lett.}, 98(23):233504, 2011.

\bibitem{Yudistira:10}
D.~Yudistira, D.~Janner, S.~Benchabane, and V.~Pruneri.
\newblock Low power consumption integrated acousto-optic filter in domain
  inverted $\mathrm{LiNbO_{3}}$ superlattice.
\newblock {\em Opt. Express}, 18(26):27181--27190, 2010.

\bibitem{6566094}
D.~Janner, D.~Tulli, M.~Jofre, D.~Yudistira, S.~Balsamo, M.~Belmonte, and
  V.~Pruneri.
\newblock Domain inverted acousto- and electrooptic devices and their
  application to optical communication, sensing, laser sources, and quantum key
  distribution.
\newblock {\em IEEE J. Sel. Topics Quantum Electron.}, 19(6):54--63, 2013.

\bibitem{yu2010fundamentals}
P.~Y. Yu and M.~Cardona.
\newblock {\em Fundamentals of semiconductors: physics and materials
  properties}.
\newblock Berlin: Springer, 2010.

\bibitem{b950586b776740e292ffa6e3bb8b2a0e}
{Elaine Cristina Saraiva} Barretto, {J{\o}rn M{\"a}rcher} Hvam, Kresten Yvind,
  and {Mike van der} Poel.
\newblock {\em Integrated Ultrasonic-Photonic Devices}.
\newblock PhD thesis, 2011.

\bibitem{narasimhamurty2012photoelastic}
T.~S. Narasimhamurty.
\newblock {\em Photoelastic and electro-optic properties of crystals}.
\newblock New York: Plenum Press, 1981.

\bibitem{PhysRevLett.87.136403}
A.~L. Ivanov and P.~B. Littlewood.
\newblock Acoustically induced stark effect for excitons in intrinsic
  semiconductors.
\newblock {\em Phys. Rev. Lett.}, 87:136403, 2001.

\bibitem{PhysRevB.66.165330}
F.~Alsina, P.~V. Santos, H.-P. Sch\"onherr, W.~Seidel, K.~H. Ploog, and
  R.~N\"otzel.
\newblock Surface-acoustic-wave-induced carrier transport in quantum wires.
\newblock {\em Phys. Rev. B}, 66:165330, 2002.

\bibitem{PhysRevLett.94.226406}
Kikuo Cho, Kazunori Okumoto, N.~I. Nikolaev, and A.~L. Ivanov.
\newblock Bragg diffraction of microcavity polaritons by a surface acoustic
  wave.
\newblock {\em Phys. Rev. Lett.}, 94:226406, 2005.

\bibitem{korpel1996acousto}
A.~Korpel.
\newblock {\em Acousto-optics}, volume~57.
\newblock New York: Marcel Dekker, 1996.

\bibitem{doi:10.1063/1.3114552}
M.~B. D\"{u}hring and O.~Sigmund.
\newblock Improving the acousto-optical interaction in a mach-zehnder
  interferometer.
\newblock {\em J. Appl. Phys.}, 105(8):083529, 2009.

\bibitem{doi:10.1063/1.2354411}
M.~M. de~Lima~Jr., M.~Beck, R.~Hey, and P.~V. Santos.
\newblock Compact $\mathrm{Mach-Zehnder}$ acousto-optic modulator.
\newblock {\em Appl. Phys. Lett.}, 89(12):121104, 2006.

\bibitem{974166}
V.~Van, T.~A. Ibrahim, K.~Ritter, P.~P. Absil, F.~G. Johnson, R.~Grover,
  J.~Goldhar, and P.~T. Ho.
\newblock All-optical nonlinear switching in gaas-algaas microring resonators.
\newblock {\em IEEE Photon. Technol. Lett.}, 14(1):74--76, 2002.

\bibitem{Wen:11}
Y.~H. Wen, O.~Kuzucu, T.~Hou, M.~Lipson, and A.~L. Gaeta.
\newblock All-optical switching of a single resonance in silicon ring
  resonators.
\newblock {\em Opt. Lett.}, 36(8):1413--1415, 2011.

\bibitem{Ksendzov:05}
A.~Ksendzov and Y.~Lin.
\newblock Integrated optics ring-resonator sensors for protein detection.
\newblock {\em Opt. Lett.}, 30(24):3344--3346, 2005.

\bibitem{5232865}
T.~Claes, J.~Girones Molera, K.~De Vos, E.~Schacht, R.~Baets, and P.~Bienstman.
\newblock Label-free biosensing with a slot-waveguide-based ring resonator in
  silicon on insulator.
\newblock {\em IEEE Photon. J.}, 1(3):197--204, 2009.

\bibitem{doi:10.1063/1.1420585}
B.~Liu, A.~Shakouri, and J.~E. Bowers.
\newblock Passive microring-resonator-coupled lasers.
\newblock {\em Appl. Phys. Lett.}, 79(22):3561--3563, 2001.

\bibitem{998697}
Bin Liu, A.~Shakouri, and J.~E. Bowers.
\newblock Wide tunable double ring resonator coupled lasers.
\newblock {\em IEEE Photon. Technol. Lett.}, 14(5):600--602, 2002.

\bibitem{rabus2007}
D.~G. Rabus~(Ed.).
\newblock {\em Integrated ring resonators}.
\newblock Berlin: Springer, 2007.

\bibitem{FAN201462}
G.~Fan, Y.~Li, C.~Hu, L.~Lei, D.~Zhao, H.~Li, and Z.~Zhen.
\newblock A novel concept of acousto-optic ring frequency shifters on
  silicon-on-insulator technology.
\newblock {\em Opt. Laser Technol.}, 63:62 -- 65, 2014.

\bibitem{fan2016process}
G.~Fan, Y.~Li, C.~Hu, L.~Lei, and Y.~Guo.
\newblock A process to control light in a micro resonator through a coupling
  modulation by surface acoustic waves.
\newblock {\em Sci. Rep.}, 6:30681, 2016.

\bibitem{tadesse2014sub}
S.~A. Tadesse and M.~Li.
\newblock Sub-optical wavelength acoustic wave modulation of integrated
  photonic resonators at microwave frequencies.
\newblock {\em Nat. Commun.}, 5:5402, 2014.

\bibitem{tadesse2016}
S.~A. Tadesse.
\newblock {\em Nano-optomechanical system based on microwave frequency surface
  acoustic waves}.
\newblock \emph{PhD Thesis} University of Minnesota, USA, 2016.

\bibitem{doi:10.1063/1.1653091}
M.~L. Dakss, L.~Kuhn, P.~F. Heidrich, and B.~A. Scott.
\newblock Grating coupler for efficient excitation of optical guided waves in
  thin films.
\newblock {\em Appl. Phys. Lett.}, 16(12):523--525, 1970.

\bibitem{Gorecki:97}
C.~Gorecki, F.~Chollet, E.~Bonnotte, and H.~Kawakatsu.
\newblock Silicon-based integrated interferometer with phase modulation driven
  by surface acoustic waves.
\newblock {\em Opt. Lett.}, 22(23):1784--1786, 1997.

\bibitem{doi:10.1117/12.281184}
C.~Gorecki, E.~Bonnotte, H.~Toshioshi, F.~Benoit, H.~Kawakatsu, and H.~Fujita.
\newblock Original approach to an optically active silicon-based
  interferometric structure for sensing applications.
\newblock volume 3098, pages 3098 -- 3098 -- 8, 1997.

\bibitem{Bonnotte:99}
E.~Bonnotte, C.~Gorecki, H.~Toshiyoshi, H.~Kawakatsu, H.~Fujita,
  K.~W\"{o}rhoff, and K.~Hashimoto.
\newblock Guided-wave acoustooptic interaction with phase modulation in a zno
  thin-film transducer on an si-based integrated mach-zehnder interferometer.
\newblock {\em J. Lightwave Technol.}, 17(1):35, 1999.

\bibitem{Crespo-Poveda:16}
A.~Crespo-Poveda, A.~Cantarero, and M.~M. de~Lima.
\newblock Reconfigurable photonic routers based on multimode interference
  couplers.
\newblock {\em J. Opt. Soc. Am. B}, 33(1):81--89, 2016.

\bibitem{doi:10.1117/12.2208868}
A.~Crespo-Poveda, A.~Hern\'{a}ndez-M\'{i}nguez, K.~Biermann, A.~Tahraoui,
  B.~Gargallo, P.~Mu\ {n}oz, P.~V. Santos, A.~Cantarero, and M.~M. de~Lima~Jr.
\newblock Tunable arrayed waveguide grating driven by surface acoustic waves.
\newblock volume 9751, pages 9751 -- 9751 -- 18, 2016.

\bibitem{1742-6596-92-1-012006}
M.~Beck, M.~M. de~Lima~Jr., R.~Hey, and P.~V. Santos.
\newblock Acoustic phonons for coherent photon control in semiconductor
  structures.
\newblock {\em J. Phys. Conf. Ser.}, 92(1):012006, 2007.

\bibitem{doi:10.1063/1.2768889}
M.~Beck, M.~M. de~Lima~Jr., E.~Wiebicke, W.~Seidel, R.~Hey, and P.~V. Santos.
\newblock Acousto-optical multiple interference switches.
\newblock {\em Appl. Phys. Lett.}, 91(6):061118, 2007.

\bibitem{doi:10.1063/1.2821306}
M.~Beck, M.~M. de~Lima~Jr., and P.~V. Santos.
\newblock Acousto-optical multiple interference devices.
\newblock {\em J. Appl. Phys.}, 103(1):014505, 2008.

\bibitem{doi:10.1117/12.853801}
E.~C.~S. Barretto and J.~M. Hvam.
\newblock Photonic integrated single-sideband modulator/frequency shifter based
  on surface acoustic waves.
\newblock volume 7719, pages 7719 -- 7719 -- 12, 2010.

\bibitem{Crespo-Poveda:13}
A.~Crespo-Poveda, R.~Hey, K.~Biermann, A.~Tahraoui, P.~V. Santos, B.~Gargallo,
  P.~Mu\ {n}oz, A.~Cantarero, and M.~M. de~Lima.
\newblock Synchronized photonic modulators driven by surface acoustic waves.
\newblock {\em Opt. Express}, 21(18):21669--21676, 2013.

\bibitem{doi:10.1117/12.2039538}
A.~Crespo-Poveda, B.~Gargallo, I.~Artundo, J.~D. Dom\'{e}nech, P.~Mu\ {n}oz,
  R.~Hey, K.~Biermann, A.~Tahraoui, P.~V. Santos, A.~Cantarero, and M.~M.
  de~Lima~Jr.
\newblock Photonic mach-zehnder modulators driven by surface acoustic waves in
  algaas technology.
\newblock volume 8989, pages 8989 -- 8989 -- 8, 2014.

\bibitem{372474}
L.~B. Soldano and E.~C.~M. Pennings.
\newblock Optical multi-mode interference devices based on self-imaging:
  principles and applications.
\newblock {\em J. Lightwave Technol.}, 13(4):615--627, 1995.

\bibitem{296191}
P.~A. Besse, M.~Bachmann, H.~Melchior, L.~B. Soldano, and M.~K. Smit.
\newblock Optical bandwidth and fabrication tolerances of multimode
  interference couplers.
\newblock {\em J. Light. Technol.}, 12(6):1004--1009, 1994.

\bibitem{crespo2016tunable}
A.~Crespo-Poveda, A.~Hern{\'a}ndez-M{\'\i}nguez, K.~Biermann, A.~Tahraoui,
  B.~Gargallo, P.~Mu{\~n}oz, P.~V. Santos, A.~Cantarero, and M.~M. de~Lima~Jr.
\newblock Tunable interferometers driven by coherent surface acoustic phonons.
\newblock {\em MRS Advances}, 1(22):1651--1656, 2016.

\bibitem{Paiam:97}
M.~R. Paiam and R.~I. MacDonald.
\newblock Design of phased-array wavelength division multiplexers using
  multimode interference couplers.
\newblock {\em Appl. Opt.}, 36(21):5097--5108, 1997.

\bibitem{Crespo-Poveda:15}
A.~Crespo-Poveda, A.~Hern\'{a}ndez-M\'{i}nguez, B.~Gargallo, K.~Biermann,
  A.~Tahraoui, P.~V. Santos, P.~Mu\ {n}oz, A.~Cantarero, and M.~M. de~Lima.
\newblock Acoustically driven arrayed waveguide grating.
\newblock {\em Opt. Express}, 23(16):21213--21231, 2015.

\bibitem{661019}
M.~Ishii, Y.~Hibino, F.~Hanawa, H.~Nakagome, and K.~Kato.
\newblock Packaging and environmental stability of thermally controlled
  arrayed-waveguide grating multiplexer module with thermoelectric device.
\newblock {\em J. Lightwave Technol.}, 16(2):258--264, 1998.

\bibitem{681462}
T.~Watanabe, N.~Ooba, S.~Hayashida, T.~Kurihara, and S.~Imamura.
\newblock Polymeric optical waveguide circuits formed using silicone resin.
\newblock {\em J. Lightwave Technol.}, 16(6):1049--1055, 1998.

\bibitem{Xiao:04}
G.~Z. Xiao, P.~Zhao, F.~G. Sun, Z.~G. Lu, Zhiyi Zhang, and C.~P. Grover.
\newblock Interrogating fiber bragg grating sensors by thermally scanning a
  demultiplexer based on arrayed waveguide gratings.
\newblock {\em Opt. Lett.}, 29(19):2222--2224, 2004.

\bibitem{5588617}
K.~Nashimoto, D.~Kudzuma, and H.~Han.
\newblock High-speed switching and filtering using $\mathrm{PLZT}$ waveguide
  devices.
\newblock In {\em OECC 2010 Technical Digest}, pages 540--542, 2010.

\bibitem{6260063}
H.~Asakura, K.~Nashimoto, D.~Kudzuma, M.~Hashimoto, and H.~Tsuda.
\newblock High-speed wavelength selective operation of $\mathrm{PLZT}$-based
  arrayed-waveguide grating.
\newblock {\em Electron. Lett.}, 48(16):1009--1010, 2012.

\bibitem{84502}
C.~Dragone.
\newblock An $\mathrm{N\times N}$ optical multiplexer using a planar
  arrangement of two star couplers.
\newblock {\em IEEE Photon. Technol. Lett.}, 3(9):812--815, 1991.

\bibitem{372441}
H.~Takahashi, K.~Oda, H.~Toba, and Y.~Inoue.
\newblock Transmission characteristics of arrayed waveguide $\mathrm{N\times
  N}$ wavelength multiplexer.
\newblock {\em J. Lightwave Technol.}, 13(3):447--455, 1995.

\bibitem{fuhrmann2011dynamic}
D.~A. Fuhrmann, S.~M. Thon, H.~Kim, D.~Bouwmeester, P.~M. Petroff, A.~Wixforth,
  and H.~J. Krenner.
\newblock Dynamic modulation of photonic crystal nanocavities using gigahertz
  acoustic phonons.
\newblock {\em Nat. Photonics}, 5(10):605, 2011.

\bibitem{fuhrmann2014high}
D.~A. Fuhrmann.
\newblock {\em High Frequency Dynamic Tuning of Photonic Crystal Nanocavities
  by Means of Acoustic Phonons}.
\newblock PhD thesis, 2014.

\bibitem{0022-3727-51-3-033001}
E.~A. Cerda-M\'{e}ndez, D.~N. Krizhanovskii, M.~S. Skolnick, and P.~V. Santos.
\newblock Quantum fluids of light in acoustic lattices.
\newblock {\em J. Phys. D: Appl. Phys.}, 51(3):033001, 2018.

\bibitem{DBMarieCuriewebsite}
https://www.sawtrain.eu/.

\end{thebibliography}

\end{document}